\documentclass[12pt,preprint]{aastex}

\shorttitle{Brightest Cluster Galaxy in 2A0335+096}
\shortauthors{Donahue et al.}
\usepackage{graphics,graphicx,epsf,natbib}
\received{2006 July 18}
\begin{document}
\title{Star Formation, Radio Sources, Cooling X-ray Gas, and Galaxy Interactions in the Brightest 
Cluster Galaxy in 2A0335+096}
\author{Megan Donahue, Ming Sun}
\affil{Michigan State University, Physics \& Astronomy Dept., East Lansing, MI 48824-2320}
\email{donahue@pa.msu.edu, sunm@pa.msu.edu}
\author{Christopher P. O'Dea}
\affil{Rochester Institute of Technology, Department of Physics, 76-3144, Rochester, NY}
\email{odea@cis.rit.edu}
\author{G. Mark Voit and Kenneth W. Cavagnolo}
\affil{Michigan State University, Physics \& Astronomy Dept., East Lansing, MI 48824-2320}
\email{voit@msu.edu, cavagnolo@pa.msu.edu}

\begin{abstract}
We present deep emission-line imaging taken with the new SOAR Optical Imaging Camera of the brightest cluster galaxy (BCG) in the nearby ($z=0.035$) X-ray 
cluster of galaxies 2A0335+096. We also present our analysis of additional,  
multi-wavelength observations for the BCG, including long-slit optical spectroscopy, 
archival VLA radio data, Chandra X-ray imaging, and XMM UV-imaging. 
Cluster 2A0335+096 is a bright, cool-core X-ray cluster, once
known as a cooling flow. Within the highly disturbed core
revealed by Chandra X-ray observations, 2A0335+096 hosts a luminous
and highly structured optical emission-line system, spanning the brightest cluster galaxy (BCG) and its companion.  
We confirm that the redshift of the companion is within 100 km s$^{-1}$ of the BCG and has 
certainly interacted with the BCG, and is likely bound to it.
The comparison of optical and radio images shows  
curved filaments in H$\alpha$ emission surrounding  
the newly resolved radio source.  The velocity structure
of the emission-line bar between the BCG nucleus and the companion galaxy
provides strong evidence for an interaction between the BCG and its northeast companion in the last $\sim50$ million years. The age of the radio source is similar to the interaction time, so this interaction may have provoked an episode of radio activity. We estimate a star formation rate of  $\gtrsim7 ~\rm{M}_\odot ~\rm{yr}^{-1}$ 
from the H$\alpha$ and archival UV data.   This rate is similar to, but somewhat lower than, the 
revised X-ray cooling rate of $10-30 ~\rm{M}_\odot ~\rm{yr}^{-1}$ in the vicinity of the BCG,  estimated from 
XMM spectra by Peterson et al. (2003). The H$\alpha$ nebula is limited to a region of  
high X-ray surface brightness and cool X-ray temperatures.  
However, the detailed structures of H$\alpha$ and X-ray gas differ.  
The peak of the X-ray surface brightness is not the peak of H$\alpha$ emission, nor does
it lie in the BCG. The estimated age of the radio lobes and their interaction with the optical emission-line gas, the estimated timescale for depletion and accumulation of cold gas, and the dynamical time in the system
are all similar, suggesting a common trigger mechanism.

\end{abstract}


\section{Introduction}

In the last few years, remarkable progress in our understanding of the properties of hot gas in the centers
of clusters of galaxies has been 
made possible by the Chandra X-ray and XMM (X-ray Multi-Mirror) 
observatories.  Stringent constraints on
emission lines from the appropriate ionization species of iron and oxygen in 
high-resolution X-ray spectra from XMM show
little evidence for simple, high $\dot{M}$ cooling flows \citep{2003ApJ...590..207P}, but indicate the presence of gas ranging over a factor of two in X-ray temperature.
High resolution imaging (to $\sim 1\arcsec$) by Chandra has revealed 
cavities in the intracluster medium (ICM),  inflated 
by radio plasma \citep{McNamaraA2597_2001,
2000ApJ...534L.135M, 2004ApJ...607..800B}. The lack of hot rims around the bubbles suggested that the
inflation has been gentle.  Chandra spectroscopic imaging has also revealed  
X-ray fronts, which are a likely to be the manifestation of cooler X-ray blobs passing through hotter gas \citep[e.g.][]{Markevitch2000A2142}.
Very deep Chandra observations have also revealed shocks emanating from the center of at least two clusters  
\citep{McNamara2005, Nulsen2005HydraA}.
Within 10 kpc of the core, structures in the hot X-ray emitting plasma 
often trace structures seen in optical emission-line images
\citep[e.g., ][]{2001MNRAS.321L..33F, McNamaraA2597_2001, Sparks2004}.

AGN feedback may be the key to stabilizing cool cores in X-ray clusters.
Such explanations have become increasingly popular to account 
for the high frequency of clusters that have a  large amount of gas with a short radiative cooling time,  multi-temperature gas, and non-radial 
spatial structures in the cluster core. The 
earliest hydrodynamic feedback 
models were produced to explain the heating and cooling of 
X-ray gas in elliptical galaxies \citep{1995MNRAS.276..663B,
1993MNRAS.263..323T}. With the Chandra discoveries of bubbles in many nearby
clusters, these models for elliptical galaxies 
evolved into feedback models for the ICM \citep[e.g., ][]{2002MNRAS.332..729C, 
2002MNRAS.331..545B, 2002ApJ...581..223R,
2002Natur.418..301B, 2002MNRAS.335..610A, 2004MNRAS.348.1105O, 2004ApJ...613..811M,
2004ApJ...615..681R,2004ApJ...617..896H,2004MNRAS.355..995D,2005ApJ...622..847S, VoitDonahue2005}
The most massive black holes in the universe, at the centers of clusters
\citep{2006astro.ph..6739L}, may produce winds, jets,
and explosions that alter the ICM entropy and prevent catastrophic cooling from occurring in clusters.

AGN feedback also appears to be an increasingly popular and important ingredient
in galaxy formation scenarios. 
Mergers may couple the growth of galaxies and their central black holes, 
thus providing a natural explanation for the unexpectedly tight relationship, discovered by \citet{Gebhardt2000} and
\citet{Ferrarese2000} between the masses of those black holes and
their host's bulge luminosity and the velocity dispersion
\citep[e.g.][]{2000MNRAS.318L..35H}. Mergers alone can not reproduce  
the exponential cutoff of the galaxy luminosity function, but mergers with AGN feedback might shut off star formation in very large galaxies \citep[e.g.][]{2006MNRAS.368L..67B,2006MNRAS.365...11C}. 
The AGN heating we now observe in low redshift systems may therefore 
give us insight into how AGN heating (and feedback) 
might have worked in the distant past.  Semi-analytic
studies currently make general assumptions about the efficiency of AGN feedback, 
tuned to suppress the formation of over-luminous, blue galaxies at the centers of clusters and
to reproduce the exponential cut-off of the galaxy luminosity function at the high end
\citep[e.g.][]{2006MNRAS.365...11C,2006MNRAS.368L..67B}. But 
these studies, because of the need for computational efficiency, 
must omit the details of how the feedback works. State-of-the-art hydrodynamic 
models of individual radio sources
generally neglect the complexity of the ambient ICM \citep[e.g.][]{2005MNRAS.357..242R,
2006AN....327..527D}. It is 
necessary, therefore, to test the 
assumptions going into these important models with detailed observations of individual
systems, in order to further the goal of making more detailed models of such systems, including triggering and termination of AGN heating.

We describe here multi-wavelength observations of the X-ray cluster 2A0335+096 that
illustrate how complex these interactions can be. 
We obtained Early Science observations from a new 4-meter class telescope (Southern Observatory for Astrophysical Research, or SOAR) and its imager (SOAR Optical Imaging Camera). 
We have combined these observations with archival 
observations of the central galaxy in order to compare 
the spatial relationship of the X-ray gas, the 
radio plasma, the emission-line gas, and the central galaxy and to estimate the star formation
rate from emission-line and UV indicators. 
We present the details of the observations, data reduction, and results in 
\S~\ref{observations}, including optical imaging from SOAR and the 
Hubble Space Telescope (\S~\ref{optical}), optical long-slit spectra from the Double Spectrograph on the
Palomar Telescope (\S~\ref{palomar}), radio images from the Very Large Array (\S~\ref{radio}), 
X-ray archival data from Chandra (\S~\ref{chandra}), and ultraviolet archival data from
the XMM Telescope (\S~\ref{om}). In \S~\ref{sect:discussion}, we discuss the implications of these 
multi-wavelength observations, beginning with 
the relationship of the brightest cluster galaxy and its companion in \S~\ref{merger}. 
We interpret the morphology of the H$\alpha$ filaments, partial rims that appear to encircle 
the resolved radio structures in \S~\ref{arcs},  and we discuss further
multi-wavelength identifications, including the correspondence between the 
X-ray, optical, and radio in \S~\ref{discussion:ids}, the dynamics of the
bar and the companion in \S~\ref{discussion:velocities}, and the star formation
and cooling rates, including UV measurements from the XMM Optical Monitor camera 
in \S~\ref{discussion:rates}. We present a summary of our results in \S~\ref{summary}.

We assume a Hubble constant of 70 km s$^{-1}$ Mpc$^{-1}$. At the redshift 
of 2A0335+096, the corresponding angular size is $0.7$ kpc arcsec$^{-1}$, and the
luminosity distance is 154 Mpc.

\section{The X-ray cluster 2A0335+096}

The X-ray cluster 2A0335+096 ($z=0.035$) is a nearby X-ray luminous
cluster with a cool core and a 
central radiative cooling time shorter than a Hubble time.
The cluster was first identified optically \citep{Zwicky1965}. 
\citet{1980ApJ...238L..59S} confirmed its positional coincidence with an 
{\em Ariel V} X-ray source \citep{1978MNRAS.182..489C} and  
noted the short central cooling time of the X-ray plasma. Its central
galaxy hosts a spectacular, filamentary emission line nebula 
\citep[][; RH88 hereafter]{RomanishinHintzen88}. 
These optical  filaments extend from a bar-shaped feature in the center of the brightest cluster
galaxy (BCG). The BCG appears to have a companion galaxy about 4 kpc projected to the NW. 
Using the {\em ROSAT HRI} detectors, 
\citet{1992ApJ...397L..31S} revealed a similar bar in the X-rays, and a larger
scale X-ray elongation of the cluster emission in the same general direction.
The BCG  is a radio galaxy  of $\sim 33.5$ mJy at 1.5 GHz diffused over a region of  
$29\arcsec \times 19\arcsec$ \citep{SBO1995}. The BCG nucleus and its companion each 
have a radio point source \citep{SBO1995}.

2A0335+096 was observed with the Chandra X-ray Observatory 
\citep{2003PASJ...55..585K,MEM2003}. We will
discuss these Chandra data further in \S~\ref{chandra}.
 The Chandra image shows a bar feature, with X-ray deficits to the NE and
the SW, identified as possible bubbles \citep{MEM2003}. But the details
of the X-ray structures and the H$\alpha$ filaments do not 
coincide, except for the bar around the nucleus, and the extension between
the BCG and the companion. 
\citet[][MEM hereafter]{MEM2003} identified a discontinuity or front feature 
in the X-ray gas to the south of the BCG, confirmed as a cold front 
by \citet[][]{2006A&A...449..475W}. MEM  
estimated a pressure jump across the feature, and derived a sub-sonic 
Mach number ($0.75\pm0.2$), or a propagation speed of 
about $400 (T/2~\rm{keV})$ km s$^{-1}$.  The two peaks in the X-ray surface brightness
map lie roughly along a line defined by the centers of the BCG and its companion. 
They discussed possible processes explaining the optical filaments in the context of their X-ray
observations, ruling out Rayleigh-Taylor instabilities, but leaving open the 
possibility that the instabilities result from a Kelvin-Helmholtz instability. 
Such instabilities could cause the H$\alpha$ filaments in a turbulent process.

Turbulence would result in broadened line profiles and, to some extent, the
lack of systematic velocity structures across the system.
Since the radial velocities of the companion and the filaments were unknown to MEM, they could
not discuss the dynamics of the interaction. Here, we  present in \S~\ref{palomar} previously unpublished 
velocity information along the axis of the interaction, redshifts
and line broadening.

\section{Observations and Data Analysis \label{observations}}
\subsection{Narrow-Band H$\alpha +$ [N~II] and Broad-band Imaging \label{optical}}

We obtained broad I-band and narrow-band images ($\Delta \lambda \sim 75$\AA, NOAO filter
ID 6781-78) centered near the redshifted H$\alpha$ line on Dec. 8, 2005 (UT) 
with the SOAR Optical Imager (SOI), mounted
on the SOAR telescope on Cerro Pachon, in Chile. We supplemented these data with
an R-band (filter F606W) Hubble Space Telescope WFPC2 image.  We used the WFPC2 association\footnote{This research utilized the facilities of the Canadian Astronomy Data Centre operated
by the National Research Council of Canada with the support of the Canadian Space 
Agency.} ID u5a40701b, which is a basic co-addition of two 300-second snapshots (Proposal ID 8301, PI Alastair Edge).

The SOAR data were obtained as part of an Early Science 
run for the SOI.  The pixels were binned $2\times2$, for a pixel scale of  
$0.154\arcsec$ per pixel.  The sky quality was excellent, with
seeing below $1\arcsec$ and good transparency. The telescope focus needed to be 
optimized on a regular basis (once every two hours) as the object transited.
We collected nine 20-minute exposures
for a total exposure time of 3 hours through the narrow band filter. Each exposure was
dithered slightly. The narrow-band filter is only 
$2$ inches $\times ~2$ inches, so the usable field of view is somewhat smaller than the full $5\arcmin \times
5\arcmin$ FOV of the SOI. After the dithering, our useful field of view resulted in
an image of $1561 \times 1395$ image pixels, or $4\arcmin \times 3.6\arcmin$.

A single 10 minute I-band exposure was obtained for the purpose of
estimating the stellar continuum contribution to the narrow-band observation.
All exposures were bias-subtracted and flat-fielded with normalized,
median-filtered twilight flats taken through the appropriate filters. 
The I-band image exhibited fringing after flat-fielding. We 
removed the fringing signal by subtracting a scaled fringe frame constructed
from dark I-band sky observations by SOAR observatory personnel.

We used the IRAF task {\em imalign} 
to align, shift, and trim the images to a common image coordinate system using five to six 
stars in the field as reference points.  The shift uncertainties were $0.04$ pixels. We then combined 
the narrow-band images by obtaining the average in each shifted pixel, after a $3\sigma$ 
clipping filter was applied. This routine removed some cosmic-ray features, but we were
able to clean single-pixel cosmic-ray events using the IRAF task {\em cosmicrays}. The pixel
scale of the SOI  is significantly smaller than the stellar PSF, so this process was reasonably successful.

The world coordinate system (WCS) telemetry from the telescope was not fully implemented at the time
of these observations, so 
we applied a post-facto astrometric WCS solution to the header of the
combined narrow-band image using tools available in 
WCSTools (v. 3.6.3, available from the Smithsonian Astrophysical Observatories). 
We matched the positions of  13 objects in the field to a catalog of USNO-A2 stars and galaxies. 
These measurements provided the central astrometric location of the  image (J2000), the angular scale of 0.154 arcseconds ($\pm0.004\arcsec$), and a very small angular offset of the field from true north.
The confidence of the absolute pointing is better than $1\arcsec$, limited by the sparse
number of actual stars in our small field. Comparisons of features in the archival F606W
HST/WFPC2 image confirmed the astrometry in the center of the field to better than $0.5\arcsec$.

\begin{figure}
\includegraphics[width=17cm]{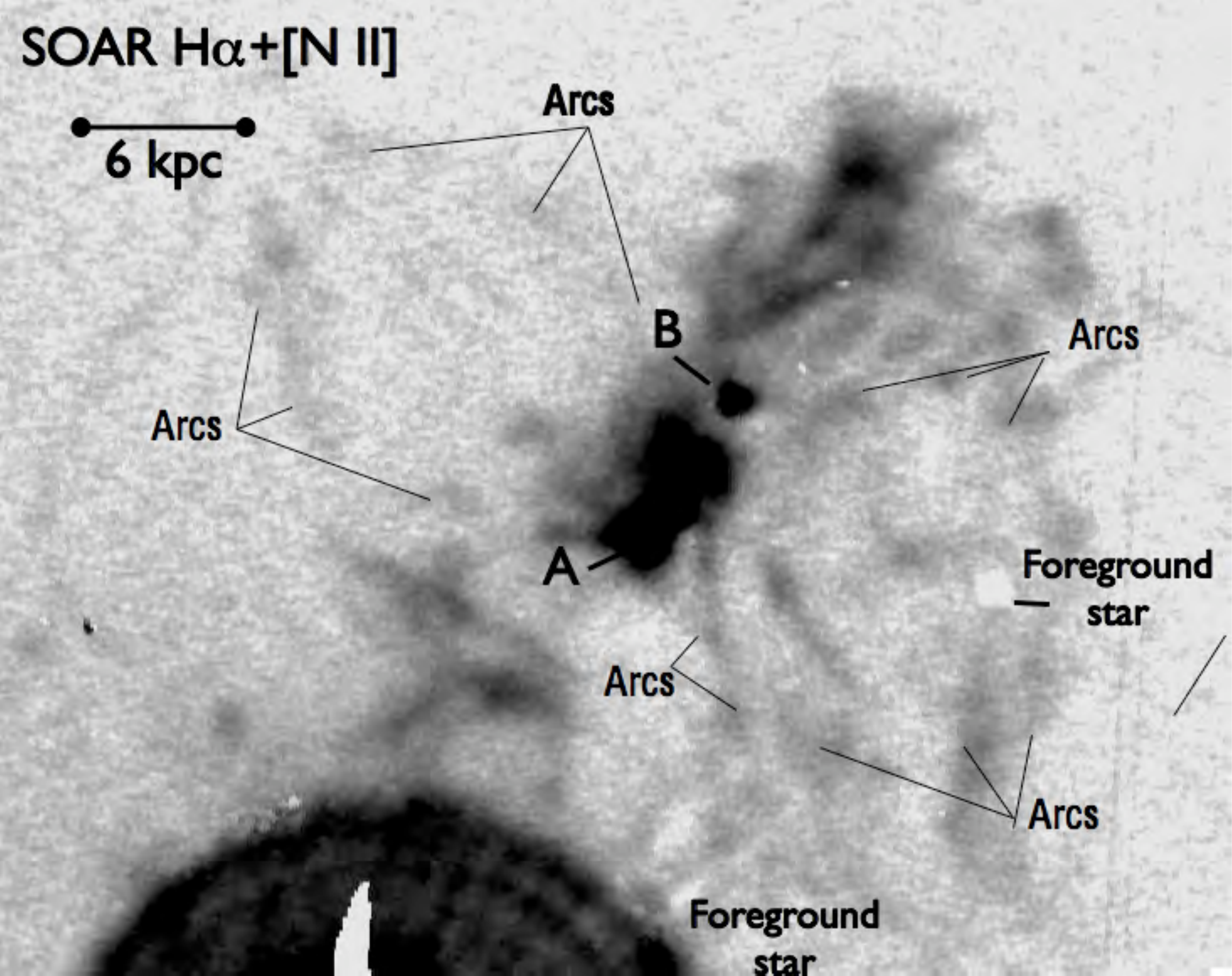}
\caption[]{A logarithmically-scaled version of the continuum-subtracted emission-line image of 
2A0335+096 from SOAR Optical Imager. North is up and East is to the left in this figure. 
The nuclei of the BCG and the companion are labeled A and B respectively. The emission-line
image shows both emission-line knots embedded in filaments. The filaments curve in arcs that
extend above and below the emission line bar. The ``arcs'' indicated here are higher 
surface brightness filaments surrounding the newly resolved radio source, and are used
to define the circles suggested in Figure~\ref{RadioContours}.
\label{soar}}
\end{figure}

\begin{figure}
\epsscale{1.5}
\includegraphics[width=17cm]{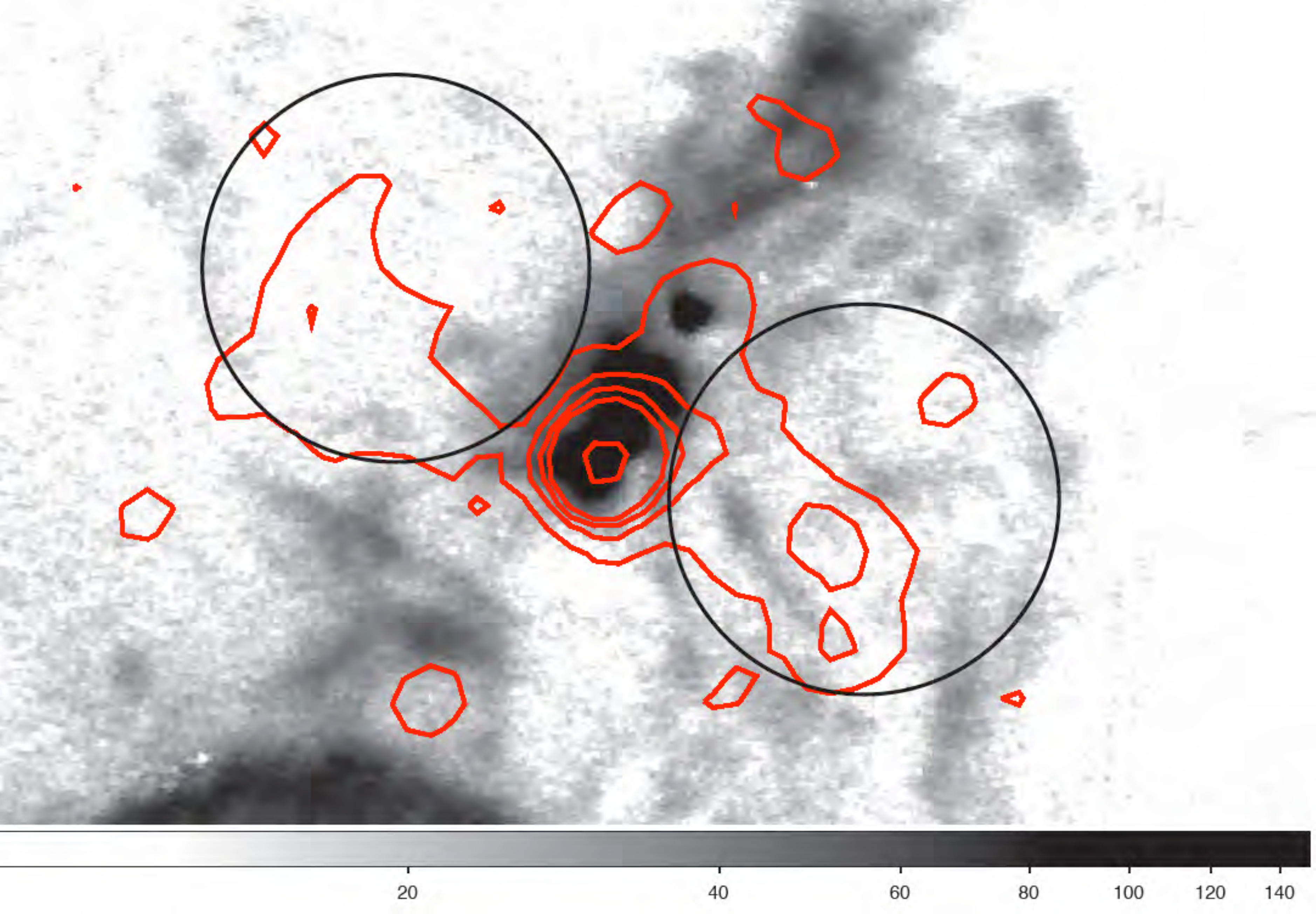}
\caption{Radio contours of the highest resolution VLA image ($3\arcsec$) 
overlaid on the optical emission line greyscale image. The radio
data were obtained Sept 22, 1987 and reprocessed on March 30, 2006.
The (red) radio contour levels are 0.5, 1.0, 1.5, 2.0, 6.0 mJy per beam.
The black circles are the locations of the putative circular 
emission-line features described in the Discussion, $8.25\arcsec$ and $8.29\arcsec$ in radius. The NE structure
is on the left, the SW structure is on the right. The
peak of the radio source aligns within $1\arcsec$ of the peak of the
emission-line source in the BCG. 
 \label{RadioContours}}
\end{figure}

Finally, we subtracted a co-aligned, fringe-corrected, scaled (by a factor of 28.57)  I-band image from the 
narrow-band image. The scale factor was chosen to subtract  
the nearby cluster galaxies from the narrow band image (Figure~\ref{soar}). 
This figure includes indicators for the filament structures used to 
visually define the circles suggested by us in Figure~\ref{RadioContours}, discussed
further in \S~\ref{radio}.
Since both [N~II] and H$\alpha$ emission features are included in the bandpass of the NOAO filter, we
use the long-slit spectroscopy described in \S3.2 to apply an approximate ($\sim15\%$
accuracy)  flux-calibration to the final net emission-line 
image of 1 ADU s$^{-1}$ per $1.26 \times 10^{-16}$ erg
s$^{-1}$ cm$^{-2}$ of pure H$\alpha$. This ratio assumes that the mean H$\alpha$ contribution
to the total H$\alpha$+[N II] emission-line flux is is 38\% throughout the nebula, based on the
long-slit spectra.

The emission-line structures are clearly visible in both the full and the
the continuum-subtracted emission-line images. No filaments or similar structures are visible in the I-band
image, which is dominated by starlight. To make a direct comparison to fluxes reported in 
RH88, we measured the total observed 
H$\alpha$ emission-line flux from the nebula within $23\arcsec$ of the BCG (masking out
a region of $r=20\arcsec$ from the bright star to the southwest)  to be 
$9 \times 10^{-14}$ erg s$^{-1}$ cm$^{-2}$.  RH88 estimate an
emission line flux of $1 \times 10^{-13}$ erg s$^{-1}$ cm$^{-2}$, if we assume the compatible
line ratios. These numbers agree well
within our estimated mutual absolute calibration accuracies of 10-15\%. Enlarging the aperture includes
a larger fraction of the detected nebula, so for the purposes of this paper, we will use
the a $35\arcsec$ aperture, for which we estimate a total pure H$\alpha$ 
flux of $(1.1 \pm 0.1) \times10^{-13}$ erg s$^{-1}$ cm$^{-2}$. The uncertainty quoted here
is based on the uncertain sky background due to sky gradients in the I-band and narrow-band
frames. An additional uncertainty arises from the scattered light contribution, arising
because the SOAR telescope was incompletely baffled at the time of the
observation. Since this scattered-light component is smooth, it does
not give rise to spurious structure in our image, but it 
does cause a large-scale gradient across
the full field ($\sim8\%$ across the twilight-flattened H$\alpha$ image). 
We note the total resolved flux from the nebula 
is almost a factor of 5 larger than that captured in the $2\arcsec \times 2\arcmin$ 
spectroscopic slit.

We note that even though the observation was quite deep, we do not detect filaments
at even larger projected distances from the BCG. A typical $3\sigma$ upper limit on such filaments,
with scales of order $1\arcsec-3\arcsec$ is about $3 \times 10^{-17}$ erg s$^{-1}$ cm$^{-2}$ arcsec$^{-2}$
of H$\alpha$ (2.6 times larger for total emission line flux from the H$\alpha$-[N~II]
complex.) 

The Galactic extinction in the direction of 2A0335+096 is $A_R=1.097$ 
\citep{1998ApJ...500..525S}. The total pure H$\alpha$ luminosity 
corrected for Galactic extinction 
but uncorrected for any internal extinction, is therefore quite large, 
$\sim 0.8 \times 10^{42}$ erg s$^{-1}$. The total star formation rate based on the H$\alpha$
luminosity, assuming the conversion of $\rm{SFR(M_\odot ~year^{-1})} =
7.9 \times 10^{-42} L(\rm{H}\alpha) ~\rm{erg ~s^{-1}}$  \citep{1998ARA&A..36..189K} 
applies in this galaxy, 
is at least 6 solar masses per year. A similar relation for starburst galaxies \citep{Kennicutt1998}
predicts at least 7 solar masses per year.

The lack of calcium emission lines 
\citep{1993ApJ...414L..17D} in the nebular gas in 
2A0335+096 indicates that calcium is depleted, and therefore that the emission-line 
gas itself is dusty.  This conclusion
is nearly independent of the details of the production of the nebular lines.
Typical dust extinction in these dusty nebula (e.g. Abell 2597;  \citet{VD1997}) is $A_V \sim 1$, so correcting for internal extinction would yield a higher 
star formation rate, up to 15-20 solar masses a year.

The H$\alpha+[N II]$ nebula in 2A0335+096 is similar in luminosity and in appearance 
to the filamentary nebula seen in Perseus's BCG (NGC1275, also 3C84) 
\citep{2003MNRAS.344L..48F}. 
Some of the emission-line structures in NGC1275 have been identified  as rising bubbles
   \citep{2003MNRAS.344L..48F, 2006MNRAS.367..433H}. The SOAR image of 2A0335+096 
shows filaments wrapping around the radiios source on both sides of the bar feature 
(Figure~\ref{RadioContours}). We will discuss these emission-line structures further in \S~\ref{arcs} 
and the radio contours in \S~\ref{radio}. The SOAR emission-line image is deeper and has
better spatial resolution than that
of RH88, revealing additional emission-line structures south and east of the bar.  However, we have detected  
no emission-line counterparts to the X-ray deficits about 38 kpc to the east and
21 kpc to the NW  of the X-ray peak (MEM).

\begin{figure}
\includegraphics[width=17cm]{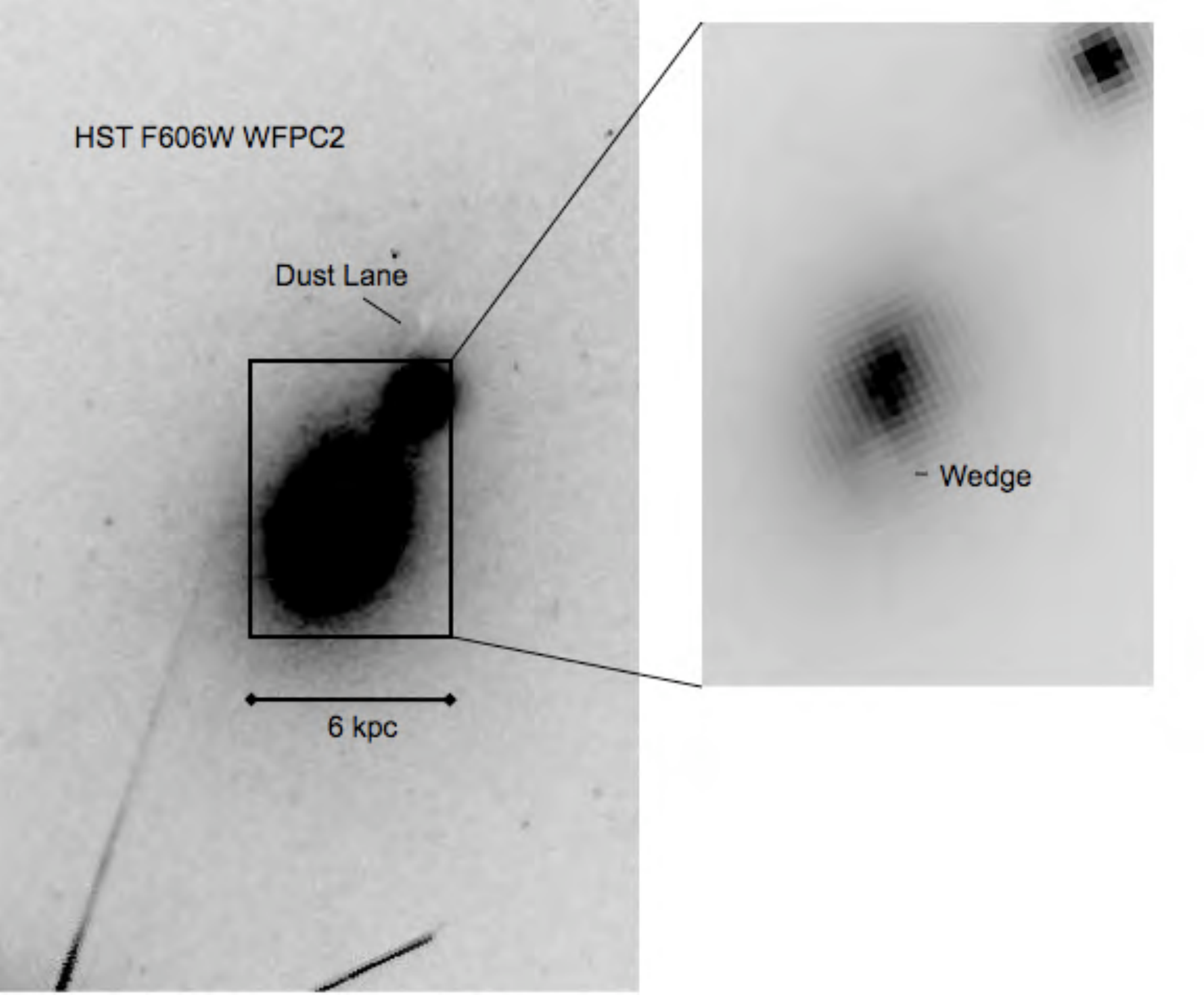}
\caption[]{A grey scale image of 2A0335+096 from the archival HST WFPC2 red (F606W) filter
imaging. North is up and East is to the left in this figure. The grey scale image to the right is 
a high contrast scaled image of the same image, indicated by the box on the left image.
\label{hst}}
\end{figure}

The HST R-band image has several notable features (Figure~\ref{hst}). The brightest galaxy and the companion have
smooth elliptical light distributions. The large-scale light distribution of the BCG
swallows the companion.  The stellar light exhibits a distinct, 
small dust lane (1-2 kpc long and $<150$ pc wide, barely resolved) 
north of the companion (Figure~\ref{hst}). This dust lane may extend somewhat 
closer to the  nucleus of the companion than is visible in the figure, and a less distinct 
N-S dust lane, $1\arcsec$ east of the companion nucleus seems to be present.
These dust lanes  appear
to be part of the general trail of material behind the companion, which suggests
that the companion may have had a tiny amount of dust in it. The only
potential dust feature visible in the BCG is  wedge-shaped,  south of the main nucleus, 
and pointing in the direction of the nucleus, as indicated in Figure~\ref{hst}. Both of these features are located in regions of H$\alpha$ emission. The wedge feature is filled in with high surface-brightness
H$\alpha$, and the northern dust lane is in the same region as a broader, similarly
aligned, H$\alpha$ filament
north of the companion.

\subsection{Long-Slit Spectroscopy \label{palomar}}

The central galaxy in 2A0335+096 was observed by Donahue and Voit 
using the 5-meter Palomar Observatory 
with the Double Spectrograph \citep{1982PASP...94..586O}  on Dec 31, 1992 (UT). 
A composite spectrum of this source was presented by \citet{1993ApJ...414L..17D}.
A $2\arcsec$ wide slit, approximately $2\arcmin$ in length was placed at a position angle of 140 degrees east of north, and centered over the galaxy and its companion.
Six 30-minute exposures were taken, with the 300-line/mm blue grating centered
on 550 nm (2.17 \AA~ per pixel) and the 1200-line/mm (0.814 
\AA~ per pixel) red grating centered on 820 nm. 
The spectral resolution was approximately 2\AA~ in the red spectrum
and 6\AA~ in the blue. The angular scale was $0.47\arcsec$ per pixel
for 24 $\mu$m pixels on the red CCD, and $0.78\arcsec$ per pixel
for $2 \times 15$ $\mu$m pixels on the blue CCD.
A composite spectrum from this observation was reported in Donahue
\& Voit (1993). The long-slit image was bias-subtracted and
dome-flattened. The wavelength calibration was obtained by fitting the known
line positions of a neon-argon arc observation and a HC observation
taken between each sky observation. Star observations taken at different
positions along the slit and the wavelength calibration were then used to
geometrically rectify each image. The sky background was removed by identifying night sky
regions on either side of the object spectrum, then fitting and  
subtracting a low-order polynomial function from the 2D image line by line.
A flux calibration for this instrumental setup was obtained from observations
of the spectroscopic 
flux standards Feige 34 and HD19445, obtained with $6\arcsec$ wide slits set to the
parallactic angle. The observations were obtained during a 50\% illuminated 
moon, but the skies were clear.  The seeing was variable and 
sometimes mediocre ($\sim 1-2\arcsec$).

\begin{figure}
\includegraphics{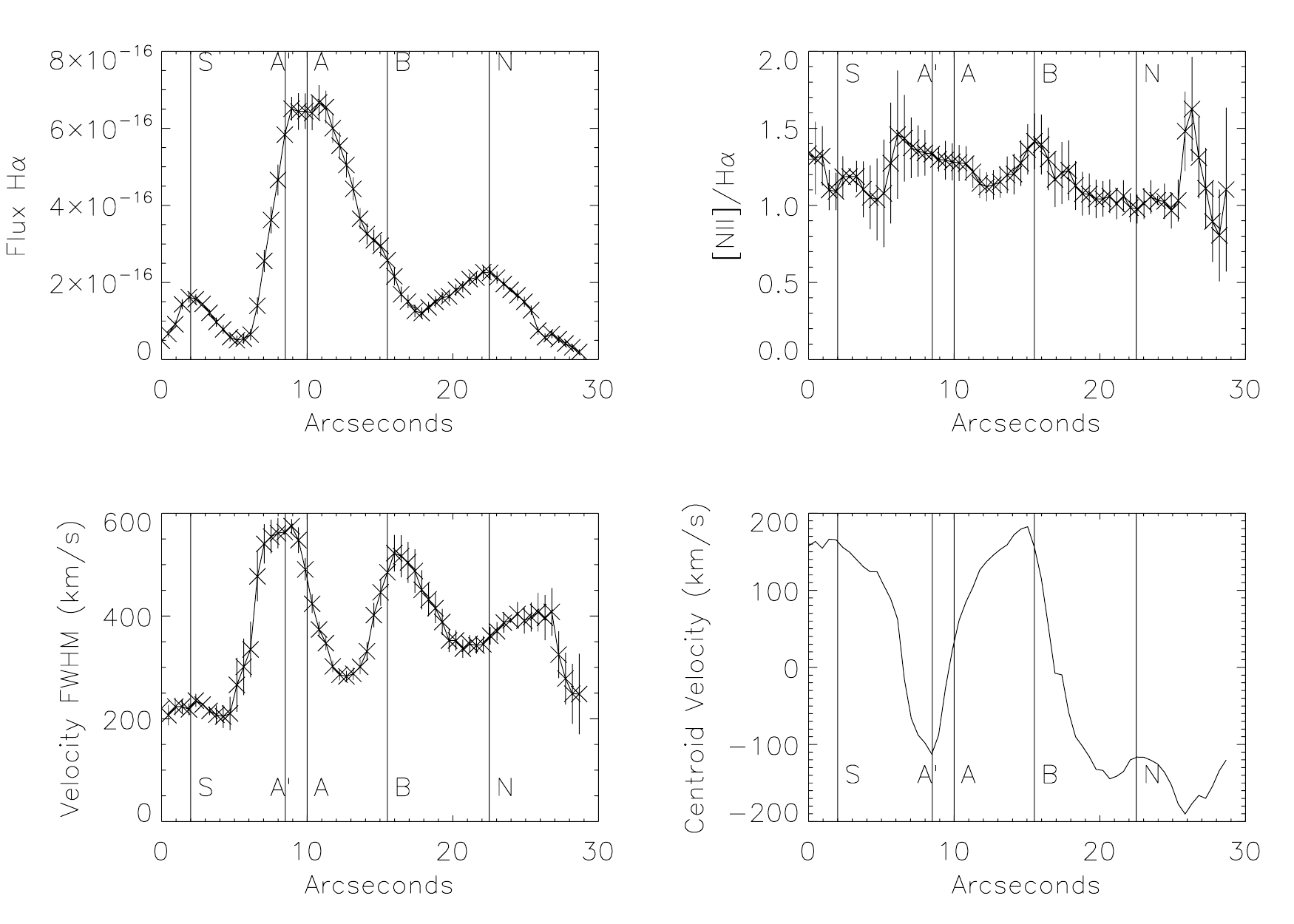}
\caption{{\bf  Upper Right.} H$\alpha$ intensity as a function of slit position. 
The brightest peak of the H$\alpha$ emission is centered in the BCG stellar continuum. 
The locations of the knots are marked by vertical lines in all the figures. Knots A and B
are the centers of the BCG and the companion. Knot A' is the location of the gas with the
broadest line profiles, located somewhat SE of the centroid of the brightest H$\alpha$ peak. 
N and S mark the location of the NW and SE gas clouds in the slit.
{\bf Lower Right.} The mean width of the H$\alpha$ and [N~II] lines in km s$^{-1}$ 
at each position along the slit. A Gaussian shape was assumed to estimate the 
FWHM from the width $\sigma$ of the Gaussian lines, where $FWHM = 2 (2 \log{2})^{1/2} \sigma$.
{\bf Upper Left.} The ratio of the integrated [N~II] 6584 line to the H$\alpha$ line, assuming that the
lines were the same width. Each line of the spectrum along the slit was fit separately. Seeing
correlates the measurements across 2-3 pixels. Three distinct peaks are seen. Two are associated
with the brightest peaks, and the third is associated with the NW filament.
{\bf Lower Left.} The mean recession velocity of all 62 positions along the slit was 
subtracted from the best-fit velocity to obtain a relative velocity for 
each slit position, in km s$^{-1}$. The mean redshift measured from these
spectra and used here is 0.0347.
\label{ha}}
\end{figure}

We have identified four major features along the spectroscopic slit, identified jointly 
in Figure~\ref{slit2d}. The main nucleus is  A, the companion nucleus is  B.
The filament (S) to the SE and the filament (N) to the NW are also labeled.
Figure~\ref{ha} plots the results of fitting each row of the 2-D 
spectrum between the wavelengths of 
6750\AA~  and 6850\AA, which includes the H$\alpha$-[N~II] emission line complex. We modeled this
complex with three redshifted, identical-width Gaussians centered at rest wavelengths of 
6548\AA, 6562.5\AA, and 6584\AA,  respectively, and a flat continuum. 
The relative emission-line strengths of the two [N~II] lines
were constrained to be in a ratio of 1:3. We fit the spectrum of each image column across the
slit. Since the pixel scale is $0.47\arcsec$ pixel$^{-1}$ and the seeing conditions varied from 
$1-2\arcsec$, these plots have effectively been passed through a 3-pixel smoothing kernel.
The approximate, relative slit positions of the image features, 
as plotted in Figures Figures~\ref{ha}  are reported in the fourth column of
Table~\ref{table:features}. 
(For reference, arcsecond numbered 0 for  Figure~\ref{ha}
corresponds to Line 63 in the original CCD frame.)

These features are distinctly seen along the slit as  peaks in the 
H$\alpha$ line intensity (Figure~\ref{ha}). The locations of these
peaks correspond approximately to changes in the  
velocity widths, relative velocity,
and [N~II]/H$\alpha$ ratio.
The emission line region A', 1.5 arcseconds SE of  
A and surrounding knot B exhibit broader emission lines (FWHM $ \sim 500-600$ km s$^{-1}$) 
than the rest of the nebula (FWHM $ \sim 200-400$ km s$^{-1}$ (Figure~\ref{ha}.) 
Since both knots are associated with radio emission, this broadening may be indicative of optical 
AGN-type activity in both nuclei (both of these nuclei are radio sources). 
The presence of bright and broadened H$\alpha$ emission is correlated with a noticeable 
enhancement of the [N~II]6584 / H$\alpha$ ratio in B, but not so much in A and A', where the associated
peak in the [N~II] ratio occurs farther south. (Figure~\ref{ha}). 
In the regions where the FWHM $> 400 $ km s$^{-1}$,
the mean [N~II]/H$\alpha$ ratio is $1.3\pm0.1$ and in the regions where the FWHM was
lower, the corresponding ratio is $1.1\pm0.15$. 
An enhanced [N~II] line relative to H$\alpha$ is suggestive of a heating source with harder
photons than typical of star formation regions \citep[e.g., ][]{BPT1981}, or a possibly a source
of supplemental heating above that from photoionization \cite[e.g., ][]{VD1997}.

An examination of the relative velocities in Figure~\ref{ha} shows the presence of at least two
distinct velocity systems  ($\bar{z} = 0.0347$), with a velocity difference of about $270$ km s$^{-1}$. The 
recession velocity of Knot A
lies (coincidentally perhaps) near the mean, in the center of a velocity gradient that extends from 
A'  (at $-120$ km/s) to B (at $+180$ km/s).
Both A' and  N are blueshifted about $-120$ to $-150$ km s$^{-1}$ relative to the mean, 
with the most extreme (and faint) velocity component of the NW  filament at $-180$ km s$^{-1}$. N
has somewhat broadened emission lines (400 km s$^{-1}$) with
elevated [N II]/H$\alpha$ as well. Both S  
and B are receding at about $+170$ km s$^{-1}$ with respect to the mean velocity of the system.

The peak of the H$\alpha$ emission along the slit is somewhat more extended 
(by a factor of two) than a point source. The south end of the peak has the broader emission lines, 
while the north end of 
the peak lies in the middle of the velocity gradient across the bar.

\subsection{Radio \label{radio}}

We re-reduced archival high-resolution 20-cm A-array data from the Very Large Array (VLA),
and added it to existing C- and D-array data. This process allowed us to produce maps at 
higher resolution than the 1.5 GHz image presented in \citet{SBO1995}. We
can see the inner lobes at a position angle of 65 degrees east of north, oriented perpendicularly to
the optical emission-line axis angle of 140-145 degrees. An extension of radio emission
towards the companion is visible along the emission-line bar. 
The higher resolution radio
map (Figure~\ref{RadioContours})  shows the radio-lobes surrounded
by curved H$\alpha$ filaments.

We estimate the age of the radiating electrons that produce the radio emission
following \citet{MyersSpangler1985}. We assume that the spectral index between
1.4 and 5 GHz has steepened from an initial value of -0.5 to the observed value
of about $-1$ due to synchrotron losses (See Table 4 of \citet{SBO1995}).
Our results are not very sensitive to whether we assume the existence of pitch angle scattering of the relativistic electrons.
Assuming the magnetic field is at the equipartition value (taking an average value for the two radio lobes of 7 $\mu$G, from 
Table 4 of \citet{SBO1995}) we estimate an age of about 25 Myr.
On the other hand, in some radio sources, there is evidence that the magnetic field is up to a factor of 4 less than the equipartition value
\citep[e.g., ][]{Carilli1994,CrostonHardcastle2005,WDW1997}. 
A factor of four weaker magnetic field implies 
an electron age of about 50 Myr.

\subsection{Chandra X-ray \label{chandra}}

We re-analyzed the Chandra data presented by MEM, in order to take advantage of the most
recent calibrations and to utilize a promising new technique developed by \citet{2006MNRAS.368..497D}
for creating X-ray surface brightness and temperature maps.
We present adaptively binned X-ray  and X-ray temperature maps of 2A 0335+096 created from the 20,000 second Chandra observation \citep{2003PASJ...55..585K} (OBSID 919). This observation was made 
with the back-illuminated Advanced CCD Imaging Spectrometer (ACIS) S3 CCD on September 6th, 2000. 
The data were prepared using  CIAO v3.3 and the calibration files in CALDB v3.2.1. Spectral fitting was performed with XSPEC 11.3.2 \citep{Arnaud1996} over the energy range $0.7-7.0$ keV.

\begin{figure}
\includegraphics[width=17cm]{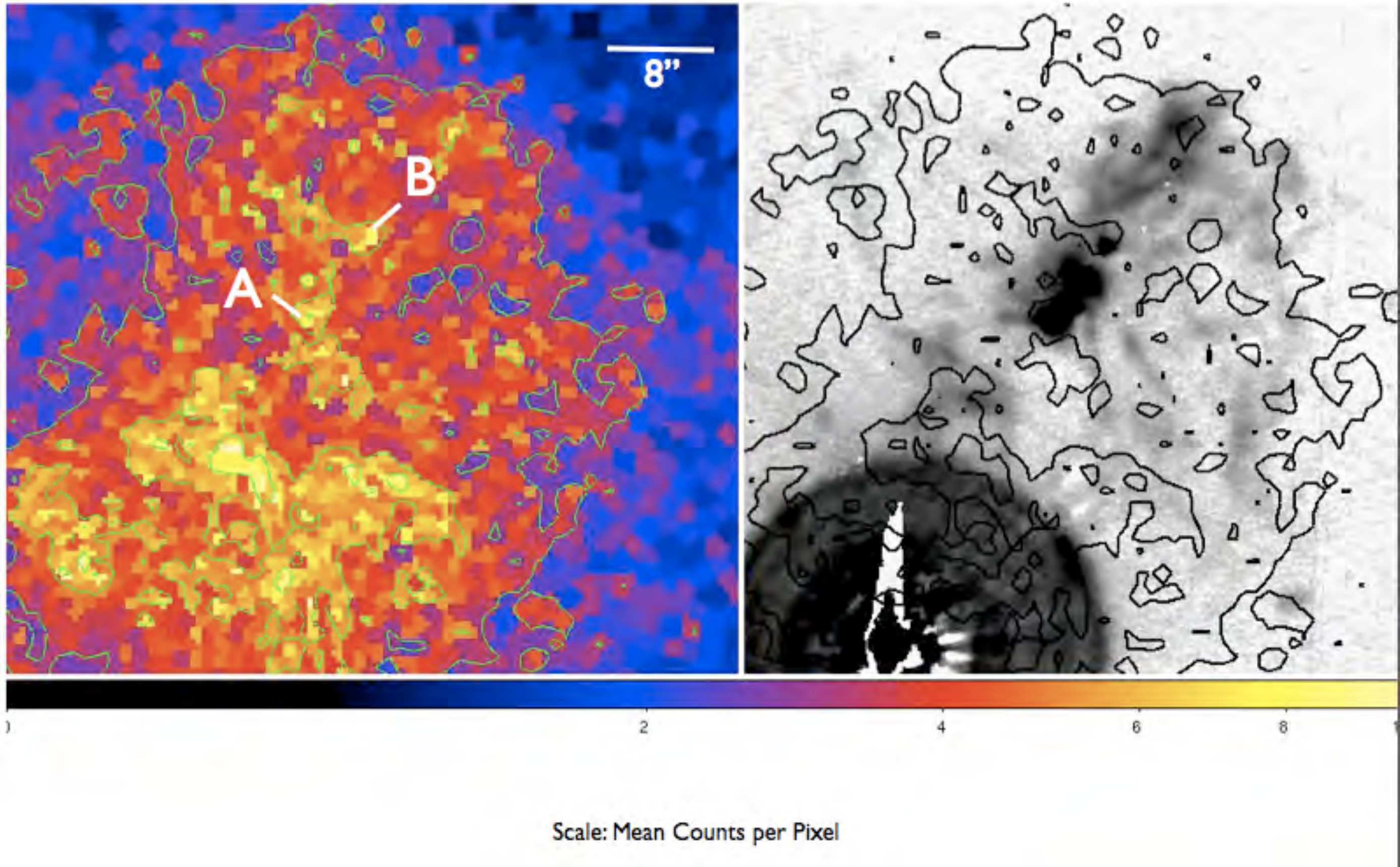}
\caption[]{Left, a color-coded image of the adaptively-binned ($5\sigma$) 
X-ray (0.7-7.0 keV) surface brightness distribution for  2A0335+096. The color bar
corresponds to the mean photon counts per ($0.5\arcsec \times 0.5\arcsec$) 
pixel in the X-ray image. The net exposure time was about 20,000 seconds.
Note that the nuclei of the galaxies A and B are not located at the brightest X-ray peak, but are 
associated with fainter X-ray peaks. Right, a contour map of the X-ray
surface brightness isophotes are plotted over the net H$\alpha$ image from SOAR, at
the same scale and orientation (North up, East left) as the left image.
While the bar of H$\alpha$ surrounding the 
BCG nucleus does appear to have a counterpart in the X-ray emission, there is
no significant one-to-one correlation of X-ray and H$\alpha$ surface brightness.  
We note that the H$\alpha$ appears to 
be limited to the region where the X-ray surface brightness is greater than about
3 photons per pixel, which is an approximate X-ray surface brightness of 
$3 \times 10^{-15}$ erg s$^{-1}$ cm$^{-2}$ arcsec$^{-2}$.

\label{xray_sb}}
\end{figure}

To create the surface brightness and temperature maps, we employed the adaptive binning code of 
\citet{2006MNRAS.368..497D}. Their technique uses weighted Voronoi tessellation (WVT) to create spatial bins which maintain a given signal-to-noise (S/N) and uses an input background spectrum specific to the observation. 
The events associated with hot pixels, bad pixels, and point sources were not included in the $S/N$ estimation for the definition of the spatial bins. Regions of the detector near the edges with less than 40\% of the total
exposure were not included in the final X-ray image.
The X-ray surface brightness map was binned to a minimum $S/N=5$ (Figure~\ref{xray_sb}).

\begin{figure}
\epsscale{.8}
\plotone{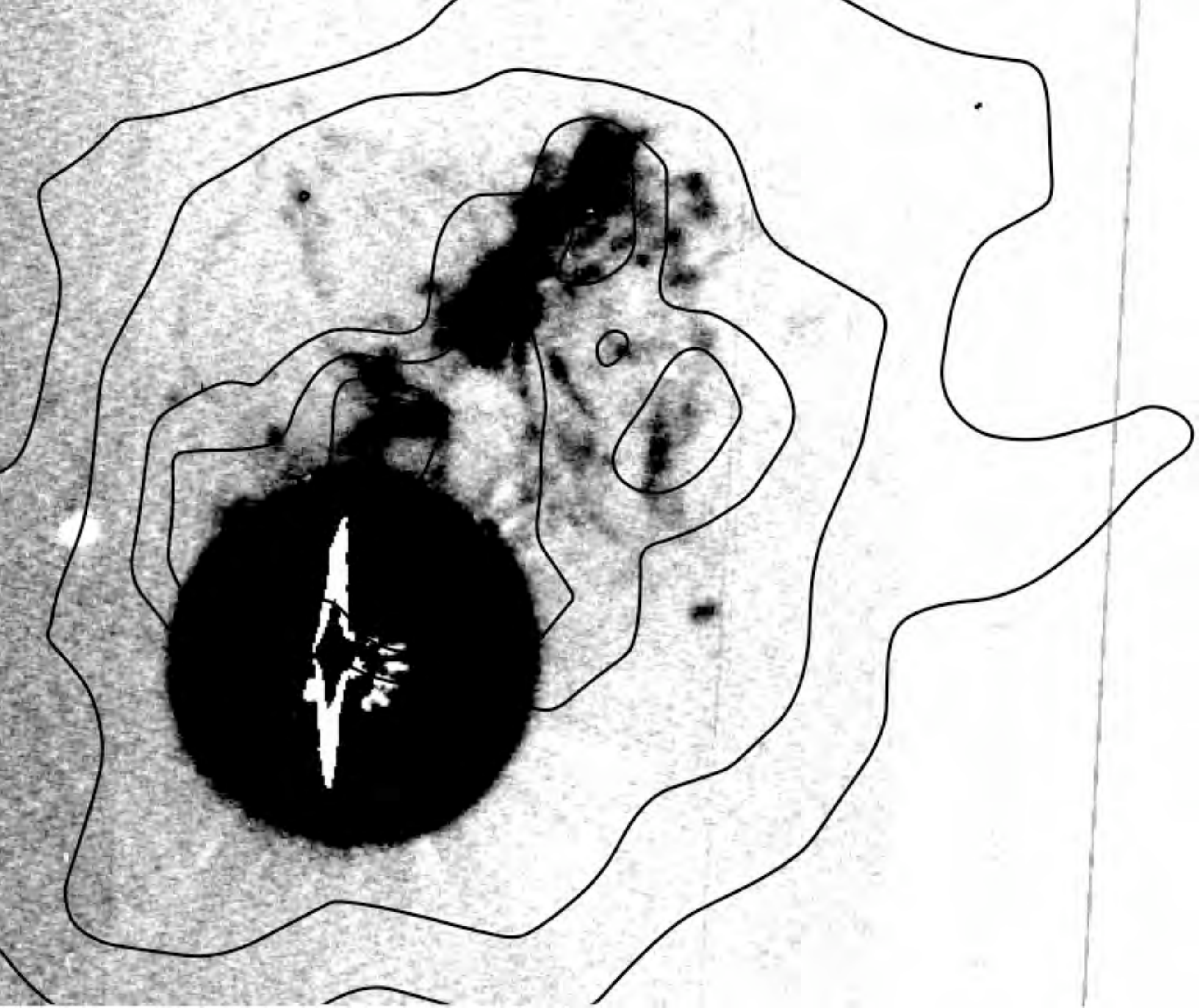}
\epsscale{1}
\caption[]{A grey-scale image of the net emission-line image, with the 
contours of the interpolated temperature map superimposed. The temperature map 
contour levels are 1.75, 2.0, 2.5, and 3.0 keV, inner to outer (the contours map a decrement
in X-ray temperatures). The H$\alpha$ filaments appear to be confined to a region where
the X-ray temperature is less than 2.5 keV, and the coolest lumps (the innermost contours, where 
$<1.75$ keV) correspond
to the brightest H$\alpha$ regions, except for the cool X-ray lump obscured by the bright
star to the south-east, which cannot be seen optically. North is up, East is left. 
\label{xray_temp}}
\end{figure}

The temperature map was extracted from 
bins defined by an X-ray surface brightness map binned to a minimum  
$S/N=30$. The high $S/N$ for the temperature map insures 
sufficient counts in each spectral bin to obtain a interesting constraint on the plasma temperature. 
To generate the temperature map, a spectrum was extracted from each WVT-defined bin. 
A corresponding background spectrum for each spatial bin was extracted from the appropriate  
deep background file (provided in the Chandra CALDB by Maxim Markevitch\footnote{ \url[]{http://cxc.harvard.edu/contrib/maxim/bg/}}), which was reprojected and matched in gain to the current observation. 
 Each spectrum was fit  with a projected single-temperature MekaL thermal
spectrum and Galactic absorption model using a fixed Galactic value of $N_{H} = 1.81 \times 10^{21}$ cm$^{-2}$.
The metallicity, temperature, and normalization 
were free parameters. The best-fit temperatures then populate a temperature map.
The best fit temperatures were also projected onto a uniform grid and interpolated using the
IDL procedures {\em triangulate} and {\em griddata}. \footnote{Interactive Data Language (IDL) is data visualization
and image analysis software available from ITT Visual Information Systems, \url{http://www.ittvis.com/}. } 
We confirmed that this interpolation did not introduce any spurious features into the temperature
map by comparing the same contour lines of both maps. The contours of the raw map oscillated around the 
smoother contours of the interpolated map, so we have chosen to show the contours of the interpolated 
temperature map  in Figure~\ref{xray_temp}, overlaid on the net H$\alpha$ image from SOAR.

\subsection{XMM/Optical Monitor: Ultraviolet \label{om}}

We reduced the {\em X-ray MultiMirror Observatory} (XMM)  
Optical/UV Monitor (OM) data of 2A0335+096, taken on August 4, 2003 (DataID 0147800201, PI
Jelle Kaastra). 
OM is a 30 cm telescope mounted on XMM that is
capable of simultaneous optical and UV observation with
the X-ray observation \citep[][]{2001A&A...365L..36M}. Exposures with
the UVW1 (2400-3600 \AA) and UVW2 (1800-2400 \AA) filters
have been taken. There were five pointings in the UVW1 band
and eight pointings in the UVW2 band, each with a central
small window (called ``Window 0" with $0.5\arcsec$ pixels) and another
bigger one (``Window 1" with $1\arcsec$ pixels \citep[][]{2001A&A...365L..36M}. 
Since the cD galaxy
is near the center of the field, it is always present  in Window 0 
of each pointing. The total UVW1 exposure is
20600 sec for  Window 0 and 4120 sec for Window 1.
The total UVW2 exposure is 34660 sec for Window 0 and
8580 sec for Window 1. The cD galaxy 2MASX J03384056+0958119
is a significant UVW1 source, as well as the companion.
There are no UVW2 detections for either source, which is not 
surprising since the telescope sensitivity in the UVW2
band is low and the known Galactic extinction in that band and 
in that direction on the sky is severe. 
Within the central $7\arcsec$ radius (enclosing
the light of the companion), the total UVW1 luminosity
is $5.5\pm0.4\times10^{42}$ erg s$^{-1}$, after a
correction for coincidence and deadtime loss. We chose
to do photometry within 7$''$ radius as that aperture
encloses most of the UV emission. The level of diffuse
UV emission beyond the aperture is highly uncertain because
of the strong scattered light near the middle of the field.
The Galactic extinction in the UVW1 band has also been
applied (2.48 mag) based on \citet[][]{Cardelli1989}, 
while the internal extinction is unknown, and is set to zero. The contribution from the
passive stellar population is subtracted from the empirical
$L_{UVW1}$ - $L_{J}$ relation derived by 
\citet[][]{2005ApJ...635L...9H}, resulting in a net NUV luminosity excess of
$2.9\pm0.5\times10^{42}$ erg s$^{-1}$. This excess corresponds
to a SFR of $3.1 - 5.7 \rm{M}_{\odot}$/yr for the assumed IMF
with a power-law index of $2.35 - 3.3$.

We have also estimated the UVW2 - UVW1 color of $>0.38$ mag,
without any correction for internal extinction. We adopt a
Galactic extinction of 3.88 mag at 2050 \AA, although there
is quite a range of extinction across the bandpass of UVW2
($3.26 - 4.14$ mag also from Cardelli et al. 1989) since the
UVW2 band encloses the peak of Galactic extinction curve
($\sim$ 2150 \AA). This color can be compared with that
of A1795's cD ($0.0 - 0.15$ mag from 
\citet[][]{2001A&A...365L..93M}. Only a small amount of internal extinction
would imply that the intrinsic color of 2A0335 is similar to that of A1795's cD.

\section{Discussion \label{sect:discussion}}

\subsection{Merger of the BCG and Companion Galaxy \label{merger}}

Most importantly, from the viewpoint of previous studies of this system, we confirm that the companion galaxy to the BCG
has a very similar recession velocity ($\Delta v \sim100$ km s$^{-1}$) to the nucleus of the BCG.
As noted by \citet[][]{2006A&A...449..475W} and by \citet[][]{MEM2003}, the cool  X-ray peak, offset from the BCG itself, 
lies in a line with the BCG
and the companion galaxy, suggesting that a subcluster has fallen in along a filament that has fed the
formation of the cluster. Our radial velocities of the companion and the BCG suggest that the companion
and the BCG are not only close in projection but may be in the process of merging..

\subsection{Emission-Line Filament Shapes and Motion of the AGN \label{arcs}}


Circles in Figure~\ref{RadioContours} show 
the positions and sizes of two apparently circular structures visible in the 
SOAR emission-line image, partially outlined by curved H$\alpha$ filaments. 
When we complete the circles outlined by these filaments, we obtained
the circles on the sky in Figure~\ref{RadioContours}.
We were struck by the symmetry and the similarity
in size of the arcs outlining curved structures in the image.
The center of the NE circle would be at $3^h~ 38^m~ 41.15^s$, 
$+09\degr 58\arcmin 19.9\arcsec$ (J2000).
The center of the SW circle would be at 
$3^h~ 38^m~ 39.8^s$, 
$+09\degr 58\arcmin 10.5\arcsec$ (J2000).
The mid-point of these two positions is 
$3^h~ 38^m~ 40.5^s$,  
$+09\degr 58\arcmin 15.1\arcsec$ (J2000).
If the AGN is at the peak of the H$\alpha$ emission, of 
$3^h~ 38^m~ 12^s$, 
$+09\degr 58\arcmin 12\arcsec$ (J2000), then the 
opening angle of the inferred structure, including the nucleus at the origin, 
is $141.5\degr$ in the plane of the sky, opening to the 
northwest, along the bar.

If the radio plasma in the lobes is 25-50 million years old, as 
we estimated in \S~\ref{radio}, 
the separation between the midpoint of the completed H$\alpha$ arcs and the actual 
radio/H$\alpha$ nucleus 
suggests that the AGN is moving through the neighboring X-ray gas at 
a velocity in the plane of the sky of about 55-110 km s$^{-1}$. This velocity is less than and therefore
consistent with 
the 400 km s$^{-1}$ 3D velocity needed to create the X-ray front feature analyzed by MEM.
The location of the extended, small-scale, radio emission along what might be the leading edge
of a cavity supports the suggestion of relative motion between the AGN and the
X-ray gas. The opening angle between the two lobes of the radio source itself is nearly straight, 
$\sim180$ degrees.

While it is possible to draw other circular structures using the filaments as guides, 
these particular structures are the
ones closest to and on opposite sides of the AGN. Therefore the velocity in the plane of the sky 
that we infer here is a lower limit. The complexity of the emission line nebula appearance indicates that
there may have been multiple outburst events, occurring along the interaction axis defined by the 
X-ray cool core, the BCG, and the companion. Multiple bubbles and a complex filamentary structure, 
reflecting a history of outbursts from
the AGN, have been seen in the X-ray image of
M87 in the center of the Virgo cluster \citep[][]{2005ApJ...635..894F}.

\subsection{Other Multi-wavelength Identifications \label{discussion:ids}}

\begin{figure}
\includegraphics[angle=0,scale=0.5]{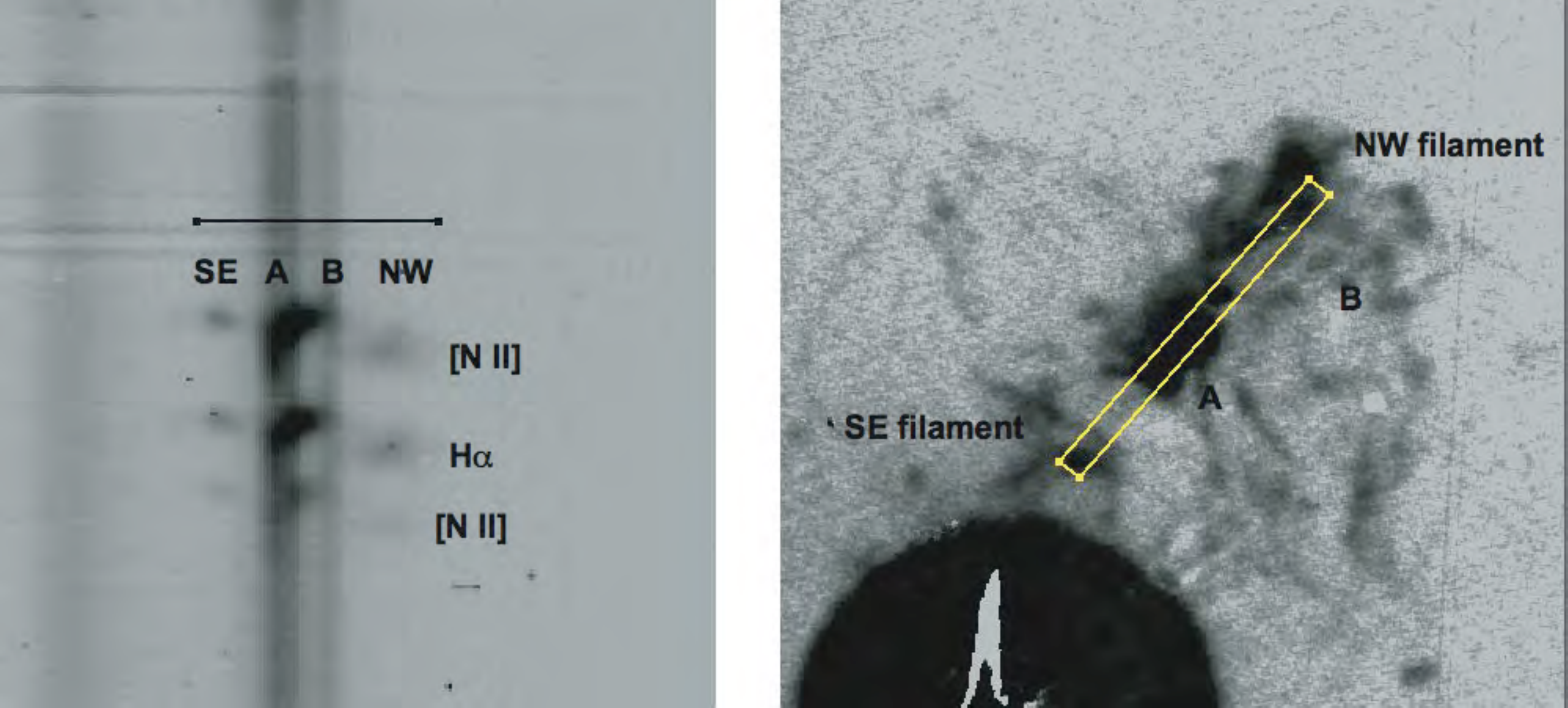}
\caption{The position angle and slit position position for the next 4 plots. 
The line on the 2-D long-slit spectrum is 60 pixels long, starting from the 
pixel 0 on the lower left hand side. The length of that line matches the length of the
slit pictured on the right-hand image showing the angle and width of the spectroscopic slit. The
actual full slit was 2 arcminutes in length and 2 arcseconds wide. 
The 60 pixels (29 arcseconds) spanning the length 
of the slit was used for the flux calibration of the SOAR emission-line image. 
\label{slit2d}}
\end{figure}

The locations of the brightest points in the BCG in the SOAR H$\alpha$ image, in the 
HST/WFPC2 image, and in the radio image are  within $0.5\arcsec$ of each
other. We identify the nucleus of the BCG with Knot A in the long-slit spectrum
(Figure~\ref{slit2d}, Figure~\ref{ha}). The 
center of the BCG is also detected in the UV by the Optical Monitor on board XMM
(\S~\ref{discussion:rates}).
The nucleus of the BCG, as seen by HST, seems to be near the center
of the brightest bar feature in the emission-line image.
The brightest peak in the X-ray emission is south of the BCG nucleus by $13.7\arcsec$ or
9.6 kpc. The BCG nucleus corresponds to a fainter peak in the X-ray emission.
Knot B (Figure~\ref{slit2d}, ~\ref{ha}) also has 
counterparts in all datasets, including the UV 
Optical Monitor (XMM), the $3\arcsec$ resolution VLA image, and
the Chandra X-ray image. 

The X-ray emission does not show a one-to-one structural correspondence with the
H$\alpha$ emission (Figure~\ref{xray_sb}), but there is a choppy ridge of 
high X-ray surface brightness that follows the more pronounced H$\alpha$ bridge between the two
galaxies.  Just outside the  H$\alpha$ nebula, to the south, and 
past the X-ray peak, there is an X-ray cool front (MEM). 
It's impossible to see the behavior of the optical emission line gas in this region  
in the SOAR image because of 
a very bright star, but even the Palomar spectrum, which is less affected by the bright star, 
does not show much hint of extended emission in that direction. 
Beyond the H$\alpha$ nebula, the X-ray surface brightness profile 
drops off suddenly, as if a minimum of X-ray surface brightness were needed to host
H$\alpha$ filaments.

Our temperature map of the cluster shows that the H$\alpha$ emission-line region
lies entirely within the 2.5 keV contour (Figure~\ref{xray_temp}). This X-ray gas has  the lowest
entropy and shortest cooling time in the cluster, and it has the same
elongated morphology as the H$\alpha$ nebula. This correspondence suggests that the
cool X-ray core and the BCG are related, even though the X-ray peak 
is not at the center of the BCG. We note that the X-ray surface brightness peaks in two
cool ($\sim1.5$ keV) clumps to the SW, one which may be associated with the SW filament and the other 
where the foreground star obscures our view of the emission-line gas. Thus the coolest X-ray
gas does not correspond exactly to the H$\alpha$ nebula, a fact we will return to in \S~\ref{discussion:rates}.
As mentioned in \S~\ref{radio} and in \S~\ref{arcs}, 
the radio source may be affecting the optical filaments on
either side of the optical/radio nucleus, as seen in the H$\alpha$ image. 
From this analysis, we concluded in \S~\ref{arcs} that the AGN 
may be in motion with respect to the optical nebula. This hypothesis is
supported by analysis of the emission-line spectra, which we discuss next.

\subsection{Dynamical Clues: Systematic Velocities along the Bar and Between the Knots
\label{discussion:velocities}}

We discuss here possible interpretations of the systematic velocity trends we
have measured along the bar of optical line emission and surrounding filaments: 
a rotating filament system in a counter-rotating disk or an interacting system,
with a bar of stripped gas. We also discuss the relationship to the motion
of the cool clump seen in the X-ray images.

This bar has been interpreted as an edge-on disk by both \citet[][]{RomanishinHintzen88} and
\citet[][]{SBO1995}. \citet[][]{SBO1995} suggest that this disk may determine the orientation of the
radio source. \citet[][]{RomanishinHintzen88} suggest a compact 2 kpc disk inside a 17 kpc bar.
Neither of these groups had spectroscopic information about the
velocities along this structure, however.
It is possible to interpret the velocity structure along the bar as 
a large, rotating filament system around Knot A 
with a circular velocity of $170$ km s$^{-1}$ (using the extreme ends of the
bar N and S to define the ``rotation"), and a diameter of 
about $20\arcsec$ or $14$ kpc. This circular velocity and size implies a 
gravitating mass of at least $5 \times 10^{10}~\rm{M}_\odot$. The center of this
system lies close to the center of the BCG.  

The inner structure of the bar could be a 
counter-rotating disk as suggested by \citet[][]{RomanishinHintzen88}, 
with a diameter of about $7.5\arcsec$ or 5.25 kpc and
a circular velocity of $\sim120-150$ km s$^{-1}$. The mean redshift of the bar may be
somewhat greater than that of the more extended system, but only by about
30 km s$^{-1}$. The intensity peak of the H$\alpha$ emission (Knot A)  lies at or near the 
center of this bar. The implied mass interior to $r=3.5$ kpc is approximately
$1.4 \times 10^{10} ~\rm{M}_\odot$. A similar, but larger-scale structure was seen in
3C275.1 \citep[][]{1986ApJ...308..540H}, but we note that even the gradient in this
structure was not considered strong indication of rotation by these authors.

A second, possibly more compelling interpretation of the velocity system is
that it is the result of an stripping interaction between the two nuclei, Knots A and B. Stripping
tails have been seen in the X-ray by \citet[][]{2006ApJ...637L..81S} in Abell 3627 and in the optical
\citep[][]{2001ApJ...563L..23G} in Abell 1367; see also Sun et al. (in preparation) in Abell 3627. 
The portion of the longslit spectrum with the broadest emission lines
seems to lie closer to the southern end of this bar, not in the center of it. 
The bar of emission
between Knots A and B may be stripped material, with a range
of velocities along the stream, stretching from Knot A towards B. If the overall direction of
motion is from NW to SE, as suggested by the orientation of the cool front and the BCG, it is
possible that the BCG, moving in the same general direction as the cool core, 
passed through B, stripping material from B in the process.

A tidal origin for the bar seems unlikely due to the lack of a similar feature in the stellar continuum
image, unless the tides also induce physical processes that light up the emission-line
and X-ray gas without significantly affecting the I-band light distribution from stars.

Knot A and the companion galaxy (Knot B)  have a velocity
separation of about 100 km s$^{-1}$ along the line of sight, and a projected separation
of 6.6$\arcsec$ or 4.6 kpc.
The crossing time therefore is several times $10^7$ years. (We have 
made the simplifying approximation that the physical separation of the two knots is $\sqrt{3}$ larger
than the projected separation, and that there is no motion in the plane of the sky.) 
  The gravitational free fall time ($t_{ff} \sim (G \rho)^{-1/2}$) 
for a system with $M\sim10^{10}~\rm{M}_\odot$ and $r=7$ kpc
is about $2 \times 10^8$ years. 

Interesting dynamical information also comes from the X-ray image. 
Recall the shape of the X-ray edge and pressure differential imply  
 that a cool clump  is  moving to the south with a Mach number of $\sim0.75$ (MEM).
This Mach number implies that the cool clump is moving 
with respect to the larger cluster at $\sim500$ km s$^{-1}$. 
The relative position of the circular H$\alpha$ filaments and the AGN 
suggest that the AGN may be moving with respect to the local ICM 
in the same direction as the
cool clump, in the plane of the sky.
The filaments defining the 
circular structure are arranged symmetrically around the radio source, 
with a slight offset suggesting lower limit to the velocity  
in the plane of the sky  of $\sim50-100$ km s$^{-1}$. These clues indicate
that the AGN may be moving in the same general direction as  the cool clump, but possibly
at a lower speed.

\subsection{Star Formation Rates and Mass Cooling Rates \label{discussion:rates}}

One problem with the original cooling flow model was the large discrepancy
between the high X-ray mass cooling rates and the low star formation rates (SFRs). 
In \S~\ref{optical}, 
we derived a minimum SFR (with no intrinsic dust extinction correction) 
of at least 7 M$_{\odot}~\rm{yr}^{-1}$ from the SOAR H$\alpha$ imaging data.
\citet{2003ApJ...590..207P} 
constrained the cooling rate of the X-ray gas in the core of
2A0335+096 with measurements and limits on X-ray emission lines seen 
in the spectrosopic grating data from the XMM RGS instrument. 
Individual X-ray lines provide independent constraints on the cooling rate,
so the cooling constraints depend on the spectral regime under consideration.
They derive an $\dot{M} (0.8 - 0.4 ~{\rm keV)}$ = 20$\pm$10
M$_{\odot}$ yr$^{-1}$ and $\dot{M} (0.4 - 0.2 ~{\rm keV)} < 84$ M$_{\odot}$ yr$^{-1}$.
While the uncertainty is large, this updated cooling rate of the X-ray gas
is close to the SFR in the cD galaxy.

From the UV observations, we derived  a near 
UV-luminosity excess of 
$2.9\pm0.5\times10^{42}$ erg s$^{-1}$. This excess corresponds
to a SFR of $3.1 - 5.7 \rm{M}_{\odot}$/yr for the assumed IMF
with a power-law index of $2.35 - 3.3$ and zero internal extinction.
If we apply an internal extinction correction 
corresponding to $A_V\sim1.0$ (typical
for cooling flow nebulae, \citep[e.g., ][]{VD1997}), the estimated SFR would be up to 10 times
larger.  The rate inferred without internal extinction is somewhat less than that inferred
from the H$\alpha$ emission, but that is to be expected since even tiny amounts of
extinction would increase the UV estimate substantially. We note the H$\alpha$ emission spans 
a much larger region ($r \sim 35\arcsec$ in radius) than that of the UV 
emission (within $7\arcsec$ radius), but that most of the H$\alpha$ is concentrated
close in. We conclude that the observed UV excess is consistent with the SFR 
inferred from the H$\alpha$ observations.

Yet another independent 
constraint on star formation was sought from the IUE data archive.
We inspected two archival IUE observations of 2A0335+096, SWP24934 and SWP43531. 
The short wavelength (SWP) camera on IUE, a 45-cm UV telescope (1978-1996) 
was sensitive to UV emission from 115-200.0 nm in a large  
aperture of $10\arcsec \times 20\arcsec$,  blueward of the XMM/OM UVW1 band. 
An upper limit to UV continuum 
was reported from the former observation of 10,800 seconds by \citet{Crawford1993}. 
We analyzed a longer, later IUE observation, 
of  24,000 seconds, but we make only a marginal improvement on Crawford et al.'s original
upper limit. Cosmic-ray events obscure the spectrum at the
position of redshifted Lyman-$\alpha$ at the nominal pointing position. A $3\sigma$ limit of $<2.4 \times 10^{-15}$
erg s$^{-1}$ cm$^{-2}$ \AA$^{-1}$, measured between 128 and 136 nm, was obtained, 
corresponding to a Galactic extinction-corrected flux limit of $6 \times 10^{-14}$ 
erg s$^{-1}$ cm$^{-2}$ \AA$^{-1}$. (No emission, extended or
otherwise, is visible in the IUE field.) A 
Lyman $\alpha$ line with a width of 500 km s$^{-1}$ would be unresolved in an 
IUE spectrum (1.67 Angstroms per element), so a $3\sigma$ line would have to 
be about $4 \times 10^{-15}$ erg s$^{-1}$ cm$^{-2}$, corresponding to an upper
limit of about $5 \times 10^{41}$ erg s$^{-1}$ on Lyman $\alpha$ luminosity and 
$<1.3 \times 10^{43}$ erg s$^{-1}$ on continuum emission in a bandpass between 128-136 nm, 
corrected for an extinction of $E(B-V)=0.41$ and a \citet{Cardelli1989}
law. 
This limit is about 50\% of what one would expect based on the H$\alpha$ luminosity
we estimate for the system with no internal extinction. However, the amount of
dust required to push the Lyman-$\alpha$ line below detectability by these IUE
data is extremely small. Therefore, the lack of detection of either a Lyman-$\alpha$ line
or an ultraviolet continuum from  2A0335+096 by IUE, together
with detections by the XMM OM and SOAR, indicates
that there is some internal dust in the BCG of 2A0335+096.


The total amount of molecular gas in the galaxy was estimated by \citet{EdgeFrayer2003}
based on CO observations
to be $1.1 \pm 0.4 \times 10^{9} h_{70}^{2}~\rm{M}_\odot$. 
 Such a gas supply, if not replenished by cooling or accretion,
would be completely used up by star formation 
in 100 million years. Therefore, this system could be classified as a starburst galaxy. 
However, since the star formation rate is similar to the estimated cooling rate from the
hot ICM by Petersen et al. (2003), we suggest that it is possible the gas supply is being
replenished by cooling.

Radio emission is thought to be related to accretion onto an AGN. This 
accretion may be fueled by a recent interaction with a gas-rich system. It is
possible that a galaxy-galaxy interaction has 
induced star formation in gas accumulated from the ICM or an ISM gas source.  
We point out that a scenario
where hot ICM has cooled and started to form stars 
is less tenable if the star-forming gas turns out to be dusty \citep{1993ApJ...414L..17D}.  
Since the dust sputtering time in 
hot gas from \citet{1979ApJ...231...77D}, the hot gas in the core of a typical cluster
sputters refractory grains with size $a<0.1 \mu$m (the majority of dust grains by number) 
in about $2 \times 10^7 n_2^{-1} a_{0.1}$ years
where $n_2=n_H/10^{-2}$ cm$^{-3}$ and $a_{0.1}=a/0.1 \mu$m, so theoretical prejudice 
is that such cooled ICM gas would be dust-free.

\section{Summary and Conclusions \label{summary}}

We summarize our results as follows.

\begin{enumerate}

\item Our SOAR H$\alpha$ image of 2A0335+096 is deeper, more
sensitive, and has better seeing  than the one presented in RH88. We clearly detect the emission-line 
filaments on 
the both the north and south sides of the galaxy, and we see filamentary details in excellent relief. 
We don't detect much more H$\alpha$ beyond $\sim25$ kpc from the central 
galaxy. These observations limit the presence of 
somewhat lower low-surface brightness emission-line gas at large radii ($\sim3 \times 10^{-17}$ 
erg s$^{-1}$ cm$^{-2}$ arcsec$^{-2}$ on scales of $1\arcsec$ - $3\arcsec$.)

\item We present velocity information along the interaction axis defined by the X-ray cool front, the
BCG, the X-ray and H$\alpha$ bar, and the companion galaxy that show the companion galaxy is
likely to have interacted with the BCG. The velocity structure along the X-ray and optical emission-line 
bar suggests that ram pressure stripping might be a relevant process here.

\item In the frame of the AGN (indicated by the center of the BCG, Knot A),
the N/NW filament/blob  is moving toward us, and 
the S/SE filament/blob is moving away from us. However, a 2-D velocity map would
be required to confirm whether the large-scale nebula is rotating in 
the frame of the AGN. We cannot tell whether the large-scale system is rotating with these data.

\item The main evidence for a earlier 
interaction between A and B comes from the morphology of
the emission bar extending between A and B, and the continuous transition of 
radial velocities from A to B seen in the long-slit spectrum.

\item Because the companion (Knot B) seems to have already
interacted with the BCG, we conclude that Knot B is now beyond the BCG, 
pulling out a bar of emission-line gas with it. Since both galaxy cores are
also emission-line sources, it is impossible to say whether B is stripping gas
from A, vice-versa, or both galaxies are stripping gas out of each other.
The passage of the companion galaxy through the BCG may have induced star formation possibly in 
pre-existing gas accumulated
from a cooling flow or in cold gas the companion may have contributed to the system. 
A stream of emission-line
gas and dust extends past the companion to the north, suggesting 
that some of the B companion's motion is also in the plane of the sky, 
and B may be now proceeding back towards the SE, towards A. 
One hint that this may be the case is that 
the gas in the NW filament has the same systematic velocity as A. Possibly ICM interactions
and gravity has decelerated that gas trailing companion B, and so its relative velocity to the
BCG is now zero.

\item The interaction may have also triggered a radio source by providing fuel or by exciting instabilities
in gas orbiting the central supermassive black hole; however, this scenario cannot be distinguished
from fueling the radio source with gas condensing from the ICM.  The higher-resolution 
radio image we present here shows two close-in lobes of radio plasma. The optical emission-line 
filaments filaments near the lobes seem to arc around the radio
plasma. The higher resolution radio data from the VLA and the deeper, higher resolution emission line imaging from SOAR 
provide complementary insight to this source.

\item We have derived a star-formation rate from the UV/OM images of 2A0335+096. These rates are 
marginally consistent with, and lower than,
constraints on the X-ray mass cooling rate from XMM RGS spectra. The approximate 
total H$\alpha$ luminosity is also consistent with the estimated star formation 
rate and rate of gas cooling from X-ray emitting
temperatures. This consistency suggests in this system at least that the star formation is a 
possible sink for gas cooling from X-ray temperatures.

\end{enumerate}

We note an interesting similarity in 
three timescales: (1)  the consumption or production timescale of the molecular gas (the 
quantity of the cold gas divided either by the X-ray cooling rate or the star formation rate), 
(2) the estimated age of the radio source derived from the 
synchrotron age of the electrons, and (3) the approximate time of last interaction between the companion and the BCG. The similarity 
in timescales suggests that the common triggering mechanism for the 
the radio source and the starburst in this system may be the interaction of a 
companion with a BCG residing in low-entropy ICM gas.

Our detailed study of this cluster core reveals details about this system that suggest the
interaction of a companion galaxy with the BCG is the trigger for the radio source and the starburst event.
Whether or not the cooling flow was required to supply the gas  that feeds the black hole and that
supplies the starburst is uncertain, but the mass cooling rates inferred from the XMM RGS spectrum
are consistent with the star formation rates inferred from H$\alpha$ and UV observations presented
here. A two-dimensional study of the emission line system, possibly with an integral field unit,
could further distinguish the dynamical models suggested by our H$\alpha$ data.


\acknowledgements

Support for this work was provided by the National Aeronautics and Space Administration (NASA) through Chandra Award Numbers SAO GO3-4159X and AR3-4017A issued by the Chandra X-ray Observatory Center, which is operated by the Smithsonian Astrophysical Observatory for and on behalf of the National Aeronautics Space Administration under contract NAS8-03060. Additional support was provided by NASA through the LTSA 
program NNG-05GD82G.
This research has made use of the NASA/IPAC Extragalactic Database (NED) which is operated by the Jet Propulsion Laboratory, California Institute of Technology, under contract with the National Aeronautics and Space Administration. We are presenting data obtained with 
the Southern Astrophysical Research Telescope, which is a joint project of Conselho Nacional de 
Pesquisas Cient'ficas e Technol—gicas CNPq-Brazil, the Univesity of North Carolina at Chapel Hill, Michigan State University, and the National Optical Astronomy Observatory.
Palomar data were obtained by MD through an agreement between the California Institute of 
Technology and the Observatories of the Carnegie Institution of Washington in Pasadena, California.
We would like to acknowledge the valuable contributions of the Palomar 
telescope operator Juan Carrasco  
and  of the SOAR service observer Dr. Kepler de Souza Oliveira.

\clearpage
\bibliography{clustercoolingflow}

\begin{thebibliography}{66}
\expandafter\ifx\csname natexlab\endcsname\relax\def\natexlab#1{#1}\fi

\bibitem[{{Alexander}(2002)}]{2002MNRAS.335..610A}
{Alexander}, P. 2002, \mnras, 335, 610

\bibitem[{{Arnaud}(1996)}]{Arnaud1996}
{Arnaud}, K.~A. 1996, in ASP Conf. Ser. 101: Astronomical Data Analysis
  Software and Systems V, ed. G.~H. {Jacoby} \& J.~{Barnes}, 17--+

\bibitem[{{B{\^ i}rzan} {et~al.}(2004){B{\^ i}rzan}, {Rafferty}, {McNamara},
  {Wise}, \& {Nulsen}}]{2004ApJ...607..800B}
{B{\^ i}rzan}, L., {Rafferty}, D.~A., {McNamara}, B.~R., {Wise}, M.~W., \&
  {Nulsen}, P.~E.~J. 2004, \apj, 607, 800

\bibitem[{{Baldwin} {et~al.}(1981){Baldwin}, {Phillips}, \&
  {Terlevich}}]{BPT1981}
{Baldwin}, J.~A., {Phillips}, M.~M., \& {Terlevich}, R. 1981, \pasp, 93, 5

\bibitem[{{Best} {et~al.}(2006){Best}, {Kaiser}, {Heckman}, \&
  {Kauffmann}}]{2006MNRAS.368L..67B}
{Best}, P.~N., {Kaiser}, C.~R., {Heckman}, T.~M., \& {Kauffmann}, G. 2006,
  \mnras, 368, L67

\bibitem[{{Binney} \& {Tabor}(1995)}]{1995MNRAS.276..663B}
{Binney}, J., \& {Tabor}, G. 1995, \mnras, 276, 663

\bibitem[{{Br{\"u}ggen} \& {Kaiser}(2002)}]{2002Natur.418..301B}
{Br{\"u}ggen}, M., \& {Kaiser}, C.~R. 2002, \nat, 418, 301

\bibitem[{{Br{\"u}ggen} {et~al.}(2002){Br{\"u}ggen}, {Kaiser}, {Churazov}, \&
  {En{\ss}lin}}]{2002MNRAS.331..545B}
{Br{\"u}ggen}, M., {Kaiser}, C.~R., {Churazov}, E., \& {En{\ss}lin}, T.~A.
  2002, \mnras, 331, 545

\bibitem[{{Cardelli} {et~al.}(1989){Cardelli}, {Clayton}, \&
  {Mathis}}]{Cardelli1989}
{Cardelli}, J.~A., {Clayton}, G.~C., \& {Mathis}, J.~S. 1989, \apj, 345, 245

\bibitem[{{Carilli} {et~al.}(1994){Carilli}, {Perley}, \&
  {Harris}}]{Carilli1994}
{Carilli}, C.~L., {Perley}, R.~A., \& {Harris}, D.~E. 1994, \mnras, 270, 173

\bibitem[{{Churazov} {et~al.}(2002){Churazov}, {Sunyaev}, {Forman}, \&
  {B{\"o}hringer}}]{2002MNRAS.332..729C}
{Churazov}, E., {Sunyaev}, R., {Forman}, W., \& {B{\"o}hringer}, H. 2002,
  \mnras, 332, 729

\bibitem[{{Cooke} {et~al.}(1978){Cooke}, {Ricketts}, {Maccacaro}, {Pye},
  {Elvis}, {Watson}, {Griffiths}, {Pounds}, {McHardy}, {Maccagni}, {Seward},
  {Page}, \& {Turner}}]{1978MNRAS.182..489C}
{Cooke}, B.~A., {Ricketts}, M.~J., {Maccacaro}, T., {Pye}, J.~P., {Elvis}, M.,
  {Watson}, M.~G., {Griffiths}, R.~E., {Pounds}, K.~A., {McHardy}, I.,
  {Maccagni}, D., {Seward}, F.~D., {Page}, C.~G., \& {Turner}, M.~J.~L. 1978,
  \mnras, 182, 489

\bibitem[{{Crawford} \& {Fabian}(1993)}]{Crawford1993}
{Crawford}, C.~S., \& {Fabian}, A.~C. 1993, \mnras, 265, 431

\bibitem[{{Croston} {et~al.}(2005){Croston}, {Hardcastle}, {Harris}, {Belsole},
  {Birkinshaw}, \& {Worrall}}]{CrostonHardcastle2005}
{Croston}, J.~H., {Hardcastle}, M.~J., {Harris}, D.~E., {Belsole}, E.,
  {Birkinshaw}, M., \& {Worrall}, D.~M. 2005, \apj, 626, 733

\bibitem[{{Croton} {et~al.}(2006){Croton}, {Springel}, {White}, {De Lucia},
  {Frenk}, {Gao}, {Jenkins}, {Kauffmann}, {Navarro}, \&
  {Yoshida}}]{2006MNRAS.365...11C}
{Croton}, D.~J., {Springel}, V., {White}, S.~D.~M., {De Lucia}, G., {Frenk},
  C.~S., {Gao}, L., {Jenkins}, A., {Kauffmann}, G., {Navarro}, J.~F., \&
  {Yoshida}, N. 2006, \mnras, 365, 11

\bibitem[{{Dalla Vecchia} {et~al.}(2004){Dalla Vecchia}, {Bower}, {Theuns},
  {Balogh}, {Mazzotta}, \& {Frenk}}]{2004MNRAS.355..995D}
{Dalla Vecchia}, C., {Bower}, R.~G., {Theuns}, T., {Balogh}, M.~L., {Mazzotta},
  P., \& {Frenk}, C.~S. 2004, \mnras, 355, 995

\bibitem[{{De Young} \& {Jones}(2006)}]{2006AN....327..527D}
{De Young}, D., \& {Jones}, T.~W. 2006, Astronomische Nachrichten, 327, 527

\bibitem[{{Diehl} \& {Statler}(2006)}]{2006MNRAS.368..497D}
{Diehl}, S., \& {Statler}, T.~S. 2006, \mnras, 368, 497

\bibitem[{{Donahue} \& {Voit}(1993)}]{1993ApJ...414L..17D}
{Donahue}, M., \& {Voit}, G.~M. 1993, \apjl, 414, L17

\bibitem[{{Draine} \& {Salpeter}(1979)}]{1979ApJ...231...77D}
{Draine}, B.~T., \& {Salpeter}, E.~E. 1979, \apj, 231, 77

\bibitem[{{Edge} \& {Frayer}(2003)}]{EdgeFrayer2003}
{Edge}, A.~C., \& {Frayer}, D.~T. 2003, \apjl, 594, L13

\bibitem[{{Fabian} {et~al.}(2003){Fabian}, {Sanders}, {Crawford}, {Conselice},
  {Gallagher}, \& {Wyse}}]{2003MNRAS.344L..48F}
{Fabian}, A.~C., {Sanders}, J.~S., {Crawford}, C.~S., {Conselice}, C.~J.,
  {Gallagher}, J.~S., \& {Wyse}, R.~F.~G. 2003, \mnras, 344, L48

\bibitem[{{Fabian} {et~al.}(2001){Fabian}, {Sanders}, {Ettori}, {Taylor},
  {Allen}, {Crawford}, {Iwasawa}, \& {Johnstone}}]{2001MNRAS.321L..33F}
{Fabian}, A.~C., {Sanders}, J.~S., {Ettori}, S., {Taylor}, G.~B., {Allen},
  S.~W., {Crawford}, C.~S., {Iwasawa}, K., \& {Johnstone}, R.~M. 2001, \mnras,
  321, L33

\bibitem[{{Ferrarese} \& {Merritt}(2000)}]{Ferrarese2000}
{Ferrarese}, L., \& {Merritt}, D. 2000, \apjl, 539, L9

\bibitem[{{Forman} {et~al.}(2005){Forman}, {Nulsen}, {Heinz}, {Owen}, {Eilek},
  {Vikhlinin}, {Markevitch}, {Kraft}, {Churazov}, \&
  {Jones}}]{2005ApJ...635..894F}
{Forman}, W., {Nulsen}, P., {Heinz}, S., {Owen}, F., {Eilek}, J., {Vikhlinin},
  A., {Markevitch}, M., {Kraft}, R., {Churazov}, E., \& {Jones}, C. 2005, \apj,
  635, 894

\bibitem[{{Gavazzi} {et~al.}(2001){Gavazzi}, {Boselli}, {Mayer},
  {Iglesias-Paramo}, {V{\'{\i}}lchez}, \& {Carrasco}}]{2001ApJ...563L..23G}
{Gavazzi}, G., {Boselli}, A., {Mayer}, L., {Iglesias-Paramo}, J.,
  {V{\'{\i}}lchez}, J.~M., \& {Carrasco}, L. 2001, \apjl, 563, L23

\bibitem[{{Gebhardt} {et~al.}(2000){Gebhardt}, {Bender}, {Bower}, {Dressler},
  {Faber}, {Filippenko}, {Green}, {Grillmair}, {Ho}, {Kormendy}, {Lauer},
  {Magorrian}, {Pinkney}, {Richstone}, \& {Tremaine}}]{Gebhardt2000}
{Gebhardt}, K., {Bender}, R., {Bower}, G., {Dressler}, A., {Faber}, S.~M.,
  {Filippenko}, A.~V., {Green}, R., {Grillmair}, C., {Ho}, L.~C., {Kormendy},
  J., {Lauer}, T.~R., {Magorrian}, J., {Pinkney}, J., {Richstone}, D., \&
  {Tremaine}, S. 2000, \apjl, 539, L13

\bibitem[{{Haehnelt} \& {Kauffmann}(2000)}]{2000MNRAS.318L..35H}
{Haehnelt}, M.~G., \& {Kauffmann}, G. 2000, \mnras, 318, L35

\bibitem[{{Hatch} {et~al.}(2006){Hatch}, {Crawford}, {Johnstone}, \&
  {Fabian}}]{2006MNRAS.367..433H}
{Hatch}, N.~A., {Crawford}, C.~S., {Johnstone}, R.~M., \& {Fabian}, A.~C. 2006,
  \mnras, 367, 433

\bibitem[{{Hicks} \& {Mushotzky}(2005)}]{2005ApJ...635L...9H}
{Hicks}, A.~K., \& {Mushotzky}, R. 2005, \apjl, 635, L9

\bibitem[{{Hintzen} \& {Stocke}(1986)}]{1986ApJ...308..540H}
{Hintzen}, P., \& {Stocke}, J. 1986, \apj, 308, 540

\bibitem[{{Hoeft} \& {Br{\"u}ggen}(2004)}]{2004ApJ...617..896H}
{Hoeft}, M., \& {Br{\"u}ggen}, M. 2004, \apj, 617, 896

\bibitem[{{Kawano} {et~al.}(2003){Kawano}, {Ohto}, \&
  {Fukazawa}}]{2003PASJ...55..585K}
{Kawano}, N., {Ohto}, A., \& {Fukazawa}, Y. 2003, \pasj, 55, 585

\bibitem[{{Kennicutt}(1998{\natexlab{a}})}]{1998ARA&A..36..189K}
{Kennicutt}, Jr., R.~C. 1998{\natexlab{a}}, \araa, 36, 189

\bibitem[{{Kennicutt}(1998{\natexlab{b}})}]{Kennicutt1998}
---. 1998{\natexlab{b}}, \apj, 498, 541

\bibitem[{{Lauer} {et~al.}(2006){Lauer}, {Faber}, {Richstone}, {Gebhardt},
  {Tremaine}, {Postman}, {Dressler}, {Aller}, {Filippenko}, {Green}, {Ho},
  {Kormendy}, {Magorrian}, \& {Pinkney}}]{2006astro.ph..6739L}
{Lauer}, T.~R., {Faber}, S.~M., {Richstone}, D., {Gebhardt}, K., {Tremaine},
  S., {Postman}, M., {Dressler}, A., {Aller}, M.~C., {Filippenko}, A.~V.,
  {Green}, R., {Ho}, L.~C., {Kormendy}, J., {Magorrian}, J., \& {Pinkney}, J.
  2006, ArXiv Astrophysics e-prints

\bibitem[{{Markevitch} {et~al.}(2000){Markevitch}, {Ponman}, {Nulsen}, {Bautz},
  {Burke}, {David}, {Davis}, {Donnelly}, {Forman}, {Jones}, {Kaastra},
  {Kellogg}, {Kim}, {Kolodziejczak}, {Mazzotta}, {Pagliaro}, {Patel}, {Van
  Speybroeck}, {Vikhlinin}, {Vrtilek}, {Wise}, \& {Zhao}}]{Markevitch2000A2142}
{Markevitch}, M., {Ponman}, T.~J., {Nulsen}, P.~E.~J., {Bautz}, M.~W., {Burke},
  D.~J., {David}, L.~P., {Davis}, D., {Donnelly}, R.~H., {Forman}, W.~R.,
  {Jones}, C., {Kaastra}, J., {Kellogg}, E., {Kim}, D.-W., {Kolodziejczak}, J.,
  {Mazzotta}, P., {Pagliaro}, A., {Patel}, S., {Van Speybroeck}, L.,
  {Vikhlinin}, A., {Vrtilek}, J., {Wise}, M., \& {Zhao}, P. 2000, \apj, 541,
  542

\bibitem[{{Mason} {et~al.}(2001){Mason}, {Breeveld}, {Much}, {Carter},
  {Cordova}, {Cropper}, {Fordham}, {Huckle}, {Ho}, {Kawakami}, {Kennea},
  {Kennedy}, {Mittaz}, {Pandel}, {Priedhorsky}, {Sasseen}, {Shirey}, {Smith},
  \& {Vreux}}]{2001A&A...365L..36M}
{Mason}, K.~O., {Breeveld}, A., {Much}, R., {Carter}, M., {Cordova}, F.~A.,
  {Cropper}, M.~S., {Fordham}, J., {Huckle}, H., {Ho}, C., {Kawakami}, H.,
  {Kennea}, J., {Kennedy}, T., {Mittaz}, J., {Pandel}, D., {Priedhorsky},
  W.~C., {Sasseen}, T., {Shirey}, R., {Smith}, P., \& {Vreux}, J.-M. 2001,
  \aap, 365, L36

\bibitem[{{Mazzotta} {et~al.}(2003){Mazzotta}, {Edge}, \&
  {Markevitch}}]{MEM2003}
{Mazzotta}, P., {Edge}, A.~C., \& {Markevitch}, M. 2003, \apj, 596, 190

\bibitem[{{McCarthy} {et~al.}(2004){McCarthy}, {Balogh}, {Babul}, {Poole}, \&
  {Horner}}]{2004ApJ...613..811M}
{McCarthy}, I.~G., {Balogh}, M.~L., {Babul}, A., {Poole}, G.~B., \& {Horner},
  D.~J. 2004, \apj, 613, 811

\bibitem[{{McNamara} {et~al.}(2005){McNamara}, {Nulsen}, {Wise}, {Rafferty},
  {Carilli}, {Sarazin}, \& {Blanton}}]{McNamara2005}
{McNamara}, B.~R., {Nulsen}, P.~E.~J., {Wise}, M.~W., {Rafferty}, D.~A.,
  {Carilli}, C., {Sarazin}, C.~L., \& {Blanton}, E.~L. 2005, \nat, 433, 45

\bibitem[{{McNamara} {et~al.}(2000){McNamara}, {Wise}, {Nulsen}, {David},
  {Sarazin}, {Bautz}, {Markevitch}, {Vikhlinin}, {Forman}, {Jones}, \&
  {Harris}}]{2000ApJ...534L.135M}
{McNamara}, B.~R., {Wise}, M., {Nulsen}, P.~E.~J., {David}, L.~P., {Sarazin},
  C.~L., {Bautz}, M., {Markevitch}, M., {Vikhlinin}, A., {Forman}, W.~R.,
  {Jones}, C., \& {Harris}, D.~E. 2000, \apjl, 534, L135

\bibitem[{{McNamara} {et~al.}(2001){McNamara}, {Wise}, {Nulsen}, {David},
  {Carilli}, {Sarazin}, {O'Dea}, {Houck}, {Donahue}, {Baum}, {Voit},
  {O'Connell}, \& {Koekemoer}}]{McNamaraA2597_2001}
{McNamara}, B.~R., {Wise}, M.~W., {Nulsen}, P.~E.~J., {David}, L.~P.,
  {Carilli}, C.~L., {Sarazin}, C.~L., {O'Dea}, C.~P., {Houck}, J., {Donahue},
  M., {Baum}, S., {Voit}, M., {O'Connell}, R.~W., \& {Koekemoer}, A. 2001,
  \apjl, 562, L149

\bibitem[{{Mittaz} {et~al.}(2001){Mittaz}, {Kaastra}, {Tamura}, {Fabian},
  {Mushotzky}, {Peterson}, {Ikebe}, {Lumb}, {Paerels}, {Stewart}, \&
  {Trudolyubov}}]{2001A&A...365L..93M}
{Mittaz}, J.~P.~D., {Kaastra}, J.~S., {Tamura}, T., {Fabian}, A.~C.,
  {Mushotzky}, R.~F., {Peterson}, J.~R., {Ikebe}, Y., {Lumb}, D.~H., {Paerels},
  F., {Stewart}, G., \& {Trudolyubov}, S. 2001, \aap, 365, L93

\bibitem[{{Myers} \& {Spangler}(1985)}]{MyersSpangler1985}
{Myers}, S.~T., \& {Spangler}, S.~R. 1985, \apj, 291, 52

\bibitem[{{Nulsen} {et~al.}(2005){Nulsen}, {McNamara}, {Wise}, \&
  {David}}]{Nulsen2005HydraA}
{Nulsen}, P.~E.~J., {McNamara}, B.~R., {Wise}, M.~W., \& {David}, L.~P. 2005,
  \apj, 628, 629

\bibitem[{{Oke} \& {Gunn}(1982)}]{1982PASP...94..586O}
{Oke}, J.~B., \& {Gunn}, J.~E. 1982, \pasp, 94, 586

\bibitem[{{Omma} {et~al.}(2004){Omma}, {Binney}, {Bryan}, \&
  {Slyz}}]{2004MNRAS.348.1105O}
{Omma}, H., {Binney}, J., {Bryan}, G., \& {Slyz}, A. 2004, \mnras, 348, 1105

\bibitem[{{Peterson} {et~al.}(2003){Peterson}, {Kahn}, {Paerels}, {Kaastra},
  {Tamura}, {Bleeker}, {Ferrigno}, \& {Jernigan}}]{2003ApJ...590..207P}
{Peterson}, J.~R., {Kahn}, S.~M., {Paerels}, F.~B.~S., {Kaastra}, J.~S.,
  {Tamura}, T., {Bleeker}, J.~A.~M., {Ferrigno}, C., \& {Jernigan}, J.~G. 2003,
  \apj, 590, 207

\bibitem[{{Reynolds} {et~al.}(2005){Reynolds}, {McKernan}, {Fabian}, {Stone},
  \& {Vernaleo}}]{2005MNRAS.357..242R}
{Reynolds}, C.~S., {McKernan}, B., {Fabian}, A.~C., {Stone}, J.~M., \&
  {Vernaleo}, J.~C. 2005, \mnras, 357, 242

\bibitem[{{Romanishin} \& {Hintzen}(1988)}]{RomanishinHintzen88}
{Romanishin}, W., \& {Hintzen}, P. 1988, \apjl, 324, L17

\bibitem[{{Roychowdhury} {et~al.}(2004){Roychowdhury}, {Ruszkowski}, {Nath}, \&
  {Begelman}}]{2004ApJ...615..681R}
{Roychowdhury}, S., {Ruszkowski}, M., {Nath}, B.~B., \& {Begelman}, M.~C. 2004,
  \apj, 615, 681

\bibitem[{{Ruszkowski} \& {Begelman}(2002)}]{2002ApJ...581..223R}
{Ruszkowski}, M., \& {Begelman}, M.~C. 2002, \apj, 581, 223

\bibitem[{{Sarazin} {et~al.}(1995){Sarazin}, {Baum}, \& {O'Dea}}]{SBO1995}
{Sarazin}, C.~L., {Baum}, S.~A., \& {O'Dea}, C.~P. 1995, \apj, 451, 125

\bibitem[{{Sarazin} {et~al.}(1992){Sarazin}, {O'Connell}, \&
  {McNamara}}]{1992ApJ...397L..31S}
{Sarazin}, C.~L., {O'Connell}, R.~W., \& {McNamara}, B.~R. 1992, \apjl, 397,
  L31

\bibitem[{{Schlegel} {et~al.}(1998){Schlegel}, {Finkbeiner}, \&
  {Davis}}]{1998ApJ...500..525S}
{Schlegel}, D.~J., {Finkbeiner}, D.~P., \& {Davis}, M. 1998, \apj, 500, 525

\bibitem[{{Schwartz} {et~al.}(1980){Schwartz}, {Schwarz}, \&
  {Tucker}}]{1980ApJ...238L..59S}
{Schwartz}, D.~A., {Schwarz}, J., \& {Tucker}, W. 1980, \apjl, 238, L59

\bibitem[{{Soker} \& {Pizzolato}(2005)}]{2005ApJ...622..847S}
{Soker}, N., \& {Pizzolato}, F. 2005, \apj, 622, 847

\bibitem[{{Sparks} {et~al.}(2004){Sparks}, {Donahue}, {Jord{\' a}n},
  {Ferrarese}, \& {C{\^ o}t{\' e}}}]{Sparks2004}
{Sparks}, W.~B., {Donahue}, M., {Jord{\' a}n}, A., {Ferrarese}, L., \& {C{\^
  o}t{\' e}}, P. 2004, \apj, 607, 294

\bibitem[{{Sun} {et~al.}(2006){Sun}, {Jones}, {Forman}, {Nulsen}, {Donahue}, \&
  {Voit}}]{2006ApJ...637L..81S}
{Sun}, M., {Jones}, C., {Forman}, W., {Nulsen}, P.~E.~J., {Donahue}, M., \&
  {Voit}, G.~M. 2006, \apjl, 637, L81

\bibitem[{{Tabor} \& {Binney}(1993)}]{1993MNRAS.263..323T}
{Tabor}, G., \& {Binney}, J. 1993, \mnras, 263, 323

\bibitem[{{Voit} \& {Donahue}(1997)}]{VD1997}
{Voit}, G.~M., \& {Donahue}, M. 1997, \apj, 486, 242

\bibitem[{{Voit} \& {Donahue}(2005)}]{VoitDonahue2005}
---. 2005, \apj, 634, 955

\bibitem[{{Wellman} {et~al.}(1997){Wellman}, {Daly}, \& {Wan}}]{WDW1997}
{Wellman}, G.~F., {Daly}, R.~A., \& {Wan}, L. 1997, \apj, 480, 79

\bibitem[{{Werner} {et~al.}(2006){Werner}, {de Plaa}, {Kaastra}, {Vink},
  {Bleeker}, {Tamura}, {Peterson}, \& {Verbunt}}]{2006A&A...449..475W}
{Werner}, N., {de Plaa}, J., {Kaastra}, J.~S., {Vink}, J., {Bleeker}, J.~A.~M.,
  {Tamura}, T., {Peterson}, J.~R., \& {Verbunt}, F. 2006, \aap, 449, 475

\bibitem[{{Zwicky} {et~al.}(1965){Zwicky}, {Karpowicz}, \&
  {Kowal}}]{Zwicky1965}
{Zwicky}, F., {Karpowicz}, M., \& {Kowal}, C.~T. 1965, in {''Catalogue of
  Galaxies and of Clusters of Galaxies'', 1965, Volume V Pasadena: California
  Institute of Technology}, 0--+

\end{thebibliography}

\clearpage

\clearpage

\begin{deluxetable}{lccccc}
\tablecaption{Summary of Emission-Line Features \label{table:features}}
\tablehead{
\colhead{Label }      & \colhead{RA (J2000)} & \colhead{Dec (J2000)}  & \colhead{Mean Slit. Pos.}\tablenotemark{a} & \colhead{Redshift } \\
 \colhead{         }       &   \colhead{               }     &  \colhead{             }         & \colhead{(arcsec)}     &  \colhead{}           \\ \tableline
}
\startdata
A                &  03:38:40.6  & +09:58:12  & 10.0    & 0.0347 (0.0343-0.0351) \\
Bridge A-B &                     &                    &         & 0.0352 (0.0352-0.0353) \\
B               & 03:38:40.3  &  +09:58:18 & 16        & 0.0353 (0.0351-0.0353)\\
NW Fil.\tablenotemark{b}   &  03:38:40    & +09:58:22  &  22.5        & 0.0343  (0.0341- 0.0347)\\
SE Fil.\tablenotemark{b}    &  03:38:41    &  +09:58:04 &  2.0          & 0.0352 (0.0351-0.0353) \\ \tableline
\enddata
\tablenotetext{a}{Each pixel in the spatial direction is 0.47 arcseconds. The slit width was 2 arcseconds.}
\tablenotetext{b}{These astrometric locations are approximate, within $3\arcsec$.}
\end{deluxetable}

\end{document}